\documentclass[preprint,floats,tightenlines,11pt,eqsecnum,showpacs,aps]{revtex4}
\usepackage{amsmath}
\usepackage{graphicx}
\begin{document}
\def\appls{\hbox{$<$\kern-.75em\lower 1.00ex\hbox{$\sim$}}}






\title{SPIN MIXING MECHANISM IN AMPLITUDE ANALYSIS OF $\pi^- p \to \pi^- \pi^+ n$\\ AND A NEW VIEW OF DARK MATTER}

\author{Miloslav Svec\footnote{electronic address: svec@hep.physics.mcgill.ca}}
\affiliation{Physics Department, Dawson College, Montreal, Quebec, Canada H3Z 1A4}
\date{December 12, 2014}

\begin{abstract}

We present the first amplitude analysis of the CERN data on $\pi^-p \to \pi^- \pi^+n$ on polarized target at 17.2 GeV/c for dipion masses 580-1080 MeV at low momentum transfers using the spin mixing mechanism. The analysis of the $S$- and $P$-wave subsystem determines a unique solution for the spin mixing transversity amplitudes $S_\tau,L_\tau$, the corresponding $S$-matrix amplitudes $S^0_\tau,L^0_\tau$ and the $\rho^0(770)-f_0(980)$ spin mixing parameters. The spin mixing mechanism allows to extract $D$-wave observables from the CERN data. Analysis of the full $D$-wave subsystem for transversity $\tau=u$ reveals $\rho^0(770)$ mixing in the amplitudes $|D^U_u|^2$ and $|D^N_u|^2$ and a violation of cosine conditions by the amplitudes $D^{2U}_u$ and $D^{2N}_u$. We determine spin mixing and $S$-matrix helicity amplitudes from which we calculate $\pi\pi$ phase-shifts $\delta^0_S$ and $\delta_P$ below $K\bar{K}$ threshold. For spin mixing amplitudes the two Solutions for $\delta^0_S$ pass through $90^\circ$ near $\rho^0(770)$ mass. There is no evidence for $\rho^0(770)-f_0(980)$ mixing in the two Solutions for $\delta^0_S$ for the $S$-matrix amplitudes. The near equality of these Solutions suggests that a unique Solution for $\delta^0_S$ is attainable in phase-shift analysis of polarized target data. 

The spin mixing and the violation of the cosine conditions arise from a non-standard pure dephasing interaction of the produced final $S$-matrix state $\rho_f(S)$ with a quantum state $\rho(E)$ of a quantum environment to produce the observed state $\rho_f(O)$. Our analysis determines that the number of interacting degrees of freedom of the environment is $M=4$. We identify the four eigenstates $|e_k>$ that define the density matrices $\rho(E)$ with the four neutrino mass eigenstes $|m_k>$ with $|m_4>$ due to light sterile neutrino. We call the mixed quantum states $\rho(E)$ dark neutrinos and propose to identify them with particles of a distinct component of dark matter. In the early Universe active neutrinos were converted in dephasing interactions into hot dark neutrinos which were redshifted by cosmic expansion to form the cold dark neutrinos of the quantum environment today. Dark neutrinos can contribute to the structure formation and evolution because their free streaming length $\lambda_{fs}(z)$ is shortened by a large number of effective degrees of freedom identified with their entropy states. With an estimated present $\lambda_{fs}(0) \sim 5$ pc or $\sim 5$ mpc they can contribute to cool or cold dark matter or even to both. Dephasing interactions involving thermal dark neutrinos and active neutrinos background are not rare events but they require high statistics accelerator experiments with polarized targets for their detection. The presented amplitude analysis illustrates this new kind of search for dark matter.

\end{abstract}

\maketitle

\tableofcontents

\newpage
\section{Introduction.}

$S$-matrix defines the unitary evolution law that evolves an initial state of particles $\rho_i$ into the final state of particles $\rho_f(S)$
\begin{equation}
\rho_f(S)=S \rho_iS^+
\end{equation}
The unitary evolution law assumes an empty Minkowski spacetime which means that  all scattering and decay processes are isolated events in the Universe. There is no environment with which the produced states $\rho_f(S)$ could interact. However we expect the physical spacetime to be permeated by various omnipresent quantum environments, such as Dark Matter, Dark Energy, Higgs field, quantum vacuum and even spacetime fluctuations. But how do we detect such quantum environment and measure its omnipresent effects on particle scattering processes that cannot be rare?

Unitary evolution law evolves pure initial states into pure final states. In 1982 Hawking pointed out that the interaction of the scattering process with the environment of spacetime fluctuations will result in a non-unitary evolution of pure initial states into mixed final states at any energy~\cite{hawking82,hawking84}. Recently we have examined the unitary evolution of pure states into pure states in $\pi N \to \pi \pi N$ processes~\cite{svec13a}. We have found that such evolution requires that the relative phases between transversity amplitudes of the same naturality and transversity must be $0$ or $\pm \pi$ in a complete disagreement with all amplitude analyses of these processes. 

The contrast between the predicted unitary relative phases and the observed non-unitary phases presents evidence for the existence of a quantum environment and its interaction with particle scattering processes~\cite{svec13a}. The interaction must be governed by a non-unitary evolution law and be consistent with the Standard Model~\cite{svec13b}. In general, the quantun state(s) of the environment have the form
\begin{equation}
\rho(E)=\sum \limits_{m,n=1}^M p_{mn}|e_m><e_n|
\end{equation}
where $|e_m>, m=1,M$ form a complete set of eigenstates describing the environment. The most general completely positive non-unitary evolution law which preserves the positivity of the probabilities is given by Kraus representation~\cite{kraus71,kraus83,nielsen00,bengtsson06,vedral06}. To be consistent with the $S$-matrix dynamics it must evolve the produced final $S$-matrix state $\rho_f(S)$ into an observed final state $\rho_f(O)$ given by
\begin{equation}
\rho_f(O)=\sum \limits_{\ell=1}^M p_{\ell \ell}V_\ell \rho_f(S)V_\ell^+
\end{equation}
where $V_\ell$ are unitary Kraus oprators. The interaction must be a pure dephasing interaction that does not change the four-momenta and the identities of the final state particles. 

The quantum environment is assumed to interact only with superpositions of diparticle spin states produced by particle production processes, such as the superpositions of dipion spins in $\pi N \to \pi \pi N$. Two-body scattering processes and free moving particles thus do not interact with the quantum environment as all Kraus operators $V_\ell$ are reduced to identity $V_\ell =I$ and $\rho_f(O)=\rho_f(S)$.

In Ref.~\cite{svec13b} we develop the non-unitary formalism and examine the consistency of the pure dephasing interaction with the conservation laws of the Standard Model in $\pi N \to \pi \pi N$ processes. In measurements with polarized targets and no measurements of recoil nucleon polarizations we work with unnatural and natural exchange transversity amplitudes $U^{J}_{\lambda,\tau}$ and $N^J_{\lambda,\tau}$, respectively~\cite{svec13a}. Here $J$ and $\lambda$ are dipion spin and helicity, and $\tau$ is the target nucleon transversity. The bilinear terms of the observed amplitudes $A(O)$ in $\rho_f(O)$ are expressed in terms of bilinear terms of so called Kraus amplitudes $A(\ell),\ell=1,M$
\begin{equation}
|A(O)||B(O)|\cos \Phi(A(O)B^*(O))=\sum \limits_{\ell=1}^M p_{\ell \ell}
|A(\ell)||B(\ell)|\cos \bigl(\phi(A(\ell))-\phi(B(\ell))\bigr)
\end{equation}
where $A$ and $B$ are unnatural or natural exchange transversity amplitudes.
There are two kinds of Kraus amplitudes. Decohering Kraus amplitudes do not mix spins and are different for each degree of freedom $\ell$. They have a general form
\begin{equation}
A(\ell)=\exp(i\alpha(\ell)) A(S)
\end{equation}
where $\exp(i\alpha(\ell))$ is a matrix element of the unitary Kraus operator $V_\ell$ and $A(S)$ is the $S$-matrix amplitude of the same type as $A(\ell)$. The phases $\alpha(\ell)$ are the dephasing phases that modify the phases of the $S$-matrix amplitudes. Decoherence free Kraus amplitudes form a special subset of Kraus amplitudes. They do not depend on the degree of freedom $\ell$ (are the same for all $\ell$) and some of them must mix spins. The decoherence free amplitudes have a general form
\begin{eqnarray}
A(\ell) & = & \exp(i\alpha) A(S)\\
B(\ell) & = & V_{BB} B(S)+V_{BC}C(S)\\
\nonumber
C(\ell) & = & V_{CB} B(S) +V_{CC}C(S)
\end{eqnarray}
where the amplitudes $A(\ell)$ do not mix spin while the amplitudes $B(\ell)$ and $C(\ell)$ mix the $S$-matrix amplitudes $B(S)$ and $C(S)$ of the same helicity $\lambda$ but different spins $J_B$ and $J_C$ such that $|J_B-J_C|=1$.  The matrix elements $V_{BB},...$ form a unitary matrix. Its elements are matrix elements of the Kraus oparator $V$ which in the decoherence free channel does not depend on $\ell$. We refer to the relation (1.7) as the spin mixing mechanism. 

In Ref.~\cite{svec13b} and in this work we focus on the amplitudes of the $S$-, $P$- and $D$-wave subsystem that dominates the pion production in $\pi^- p \to \pi^- \pi^+ n$ and $\pi^+ n \to \pi^+ \pi^- p$ below 1400 MeV. The amplitudes $S_\tau$, $L_\tau$ and $D^0_\tau$ are the $S$-,$P$- and $D$-wave helicity $\lambda=0$ unnatural exchange amplitudes. The amplitudes $U_\tau$ and $N_\tau$ are the unnatural and natural exchange $P$-wave amplitudes with helicity $\lambda=1$. $D$-wave ampitudes $D^U_\tau$ and $D^{2U}_\tau$ are unnatural exchange amplitudes with helicitites $\lambda=1$ and $\lambda=2$, respectively. Similarly, the amplitudes $D^N_\tau$ and $D^{2N}_\tau$ are the $D$-wave natural exchange ampitudes with $\lambda=1,2$.

The consistency of the pure dephasing interaction with the Standard Model predicts theoretically a spin mixing of two pairs of amplitudes: $S_\tau$, $L_\tau$ and $U_\tau$, $D^U_\tau$~\cite{svec13b}. The first unrecognized evidence for $\rho^0(770)-f_0(980)$ spin mixing in $S_\tau$, $L_\tau$ amplitudes dates back to 1960's~\cite{hagopian63,islam64,patil64,durand65,baton65,donohue79} and was later confirmed in amplitude analyses of measurements on polarized targets of $\pi^- p \to \pi^- \pi^+  n$ at 17.2GeV/c~\cite{becker79a,becker79b,chabaud83,sakrejda84,rybicki85,kaminski97,kaminski02,svec92c,svec96,svec97a,svec12a} and at 1.78 GeV/c~\cite{alekseev99} as well as of $\pi^+ n \to \pi^+ \pi^- p$ at 5.98 and 11.85 GeV/c~\cite{svec92c,svec96,svec97a}. A survey of experimenal evidence for $\rho^0(770)-f_0(980)$ spin mixing from all amplitude analyses of the five measurements on polarized targets is given in Ref~\cite{svec12d}.

Based on experimental evidence the amplitudes $S_\tau(\ell),
L_\tau(\ell),U_\tau(\ell),N_\tau(\ell),D^0_\tau(\ell)$ form decoherence free subsystem. The spin mixing mechanism for the amplitudes $S_\tau$ and $L_\tau$ reads~\cite{svec13b}
\begin{eqnarray}
L_\tau (\ell) & = & e^{i\phi} \bigl ( +\cos \theta S^0_\tau + e^{i\phi} \sin\theta L^0_\tau \bigr )\\
\nonumber
S_\tau (\ell) & = & e^{i\phi} \bigl ( -\sin \theta S^0_\tau + e^{i\phi} \cos\theta L^0_\tau \bigr )
\end{eqnarray}
where $\phi$ and $\theta$ are the spin mixing parameters and where we use the superscript $0$ to label the $S$-matrix amplitudes. We assume the following decoherence free amplitudes do not mix spins~\cite{svec13b}
\begin{eqnarray}
U_\tau (\ell) & = & e^{i2\phi} U^0_\tau\\
\nonumber
N_\tau (\ell) & = & e^{i2\phi} N^0_\tau\\
\nonumber
D^0_\tau (\ell) & = & e^{i\psi} D^{0,0}_\tau
\end{eqnarray}
In general $U_\tau$ and $N_\tau$ and their dephasing decoherence free partners $D^U(\ell)$ and $D^N(\ell)$ will mix spins~\cite{svec13b}. The remaining $D$-wave amplitudes are decohering amplitudes
\begin{eqnarray}
D^{2U}_\tau(\ell) & = & \exp(i\chi(\ell))D^{2U,0}_\tau\\
\nonumber
D^{2N}_\tau(\ell) & = & \exp(i\chi(\ell))D^{2N,0}_\tau
\end{eqnarray}
where $\ell=1,M$ and $2 \leq M \leq 4$~\cite{svec13b}.

In this work we present a new amplitude analysis of the CERN measurements of $\pi^-p \to \pi^- \pi^+ n$ on polarized target at 17.2 GeV/c for dipion masses 580 - 1080 MeV and low $|t|\leq 0.20$(GeV/c)$^2$. The first objective of the analysis is to determine the spin mixing parameters $\phi$, $\theta$ and the $S$-matrix amplitudes $|S^0_\tau|^2$ and $|L^0_\tau|^2$. The second objective is to determine the moduli of the $D$-wave amplitudes, their phases $\psi$, $\eta(\ell)$ and $\chi(\ell)$, and to determine the dimension $M$.

The starting point of the analysis are two groups of measured observables corresponding to target transversities $\tau=u$ (spin "up") and $\tau=d$ (spin "down"). Previous analyses established that the $S$- and $P$-wave moduli for $\tau=d$ are about three times larger than those for $\tau=u$ which allows us to neglect the $D$-wave contributions in the system of equations "down". In this $S$- and $P$-wave system we then use the spin mixing mechanism (SMM) to solve for the spin mixing parameters and the $S$-matrix amplitudes $S^0_d$ and $L^0_d$. We find a single physical solution. Next we use these resuts, SMM and some "up" observables to determine the $S$- and $P$-wave observables for $\tau=u$. The difference between the measured "up" observables and the corresponding $S$- and $P$-wave "up" observables allows us to determine the moduli and phases of the $D$-wave amplitudes with the use of certain enabling assumptions. We find evidence for $\rho^0(770)$ mixing in the amplitudes $D^U$ and $D^N$ and that $M=4$. 

Helicity amplitudes allow us to calculate the $\pi \pi$ phase-shifts $\delta^0_S$ and $\delta_P$ below the $K\bar{K}$ threshold. There are two Solutions for $\delta^0_S$. For the spin mixing helicity amplitudes both Solutions pass through $90^\circ$ near the $\rho^0(770)$ mass. Apart from the mass region 830-930 MeV, the Solution 1 is similar to Solution "down-flat" from the Cracow phase-shift analysis~\cite{kaminski97} while the Solution 2 is similar to their Solution "down-steep". For the $S$-matrix helicity amplitudes both Solutions for $\delta^0_S$ are essentially flat with no evidence of $\rho^0(770)-f_0(980)$ mixing. The near equality of these two Solutions suggests that a unique Solution for the phase-shift $\delta^0_S$ consistent with $S$-matrix unitarity is attainable from the data on polarized targets.

The paper is organized as follows. In Section II. we define the observables measured in $\pi N \to \pi \pi N$ on polarized targets. In Section III. we present expressions of the observables for the $S$-, $P$- and $D$-wave subsystem in terms of the bilinear terms of transversity amplitudes. In Section IV. we clarify the relationship between the observed bilinear terms and the bilinear terms of Kraus amplitudes. In Section V. we present our new amplitude analysis of the $S$- and $P$-wave subsystem using the spin mixing mechanism. In Section VI. we present our amplitude analysis of the $D$-wave subsystem. In Section VII. we determine the helicity amplitudes and in Section VIII. we calculate the phase shifts $\delta^0_S$ and $\delta_P$ and compare our Solutions to the Cracow results.

In Section IX. we present a new view of dark matter. We propose that the four eigenstates $|e_i>$ that form the quantum states $\rho(E)$ of the environment are the four mass neutrino eigenstates $|m_i>$ including light sterile neutrino $|m_4>$. While active neutrinos are pure states, the states $\rho(E)$ are mixed states which we call "dark neutrinos". We identify these dark neutrinos with particles of a distinct component of dark matter. Hot dark neutrinos were created in pure dephasing interactions of active neutrinos with particle scattering processes in the early Universe and were redshifted to form late cold dark neutrinos of cold dark matter. Dark neutrinos can contribute to the structure formation because their free streaming length is shortened by the large number of their entropy states. The dephasing interactions involving cold dark neutrinos and active neutrinos background are not rare events but require high statistics accelerator measurements with polarized targets for their detection. The paper closes in Section X. with a summary and a discussion.

\section{The observables in $\pi N \to \pi \pi N$ on polarized target.}

We consider the pion creation process $\pi N \to \pi \pi N$ with four-momenta $p_a+p_b = p_1+p_2+p_d$. In the laboratory system of the reaction the $+z$ axis has the direction opposite to the incident pion beam. The $+y$ axis is perpendicular to the scattering plane and has direction of $\vec {p}_a \times \vec {p}_c$ where $p_c=p_1+p_2$. The angular distribution of the produced dipion system is described by the direction of $\pi^-$ in the two-pion center-of-mass system and its solid angle $\Omega = \theta, \phi$. The target polarization vector is $\vec{P}=(P_x,P_y,P_z)=(P_T\sin \psi,P_T\cos \psi, P_L)$ where $P_T$ and $P_L$ are transverse and logitudinal polarization components perpendicular and parallel to the $z$-axis, respectively, and $\psi$ is the angle between $\vec{P_T}$ and the $y$-axis. The invariant mass of the dipion system is $m^2=(p_1+p_2)^2$.

When the polarization of the recoil nucleon is not measured the angular intensity takes the form~\cite{svec13a,lutz78,sakrejda84}
\begin{equation}
I(\Omega,\psi)=I_U(\Omega)+P_T\cos \psi I_C(\Omega)+P_T\sin \psi I_S(\Omega)+P_L I_L(\Omega)
\end{equation}
We shall use the parametrization of the angular components $I_U,I_C,I_S,I_L$ due to Lutz and Rybicki~\cite{lutz78,becker79a,becker79b,chabaud83,sakrejda84,rybicki85}
\begin{eqnarray}
I_U(\Omega) & = & \sum \limits_{L,M} t^L_M ReY^L_M(\Omega)\\
\nonumber
I_C(\Omega) & = & \sum \limits_{L,M} p^L_M ReY^L_M(\Omega)\\
\nonumber
I_S(\Omega) & = & \sum \limits_{L,M} r^L_M ImY^L_M(\Omega)\\
\nonumber
I_L(\Omega) & = & \sum \limits_{L,M} q^L_M ImY^L_M(\Omega)
\end{eqnarray}
The parametrization (2.2) assumes $P$-parity conservation. The parameters $t,p,r,q$ are related to the moments of angular distributions used in Ref.~\cite{grayer74}
\begin{eqnarray}
<ReY^L_M> = \frac{1}{2\pi} \int I(\Omega)ReY^L_M(\Omega d\Omega d\psi & = & \frac{1}{E_M}t^L_M\\
\nonumber
<\cos \psi ReY^L_M> = \frac{1}{2\pi} \int I(\Omega)\cos \psi ReY^L_M(\Omega d\Omega d\psi & = & \frac{1}{2E_M}p^L_M\\
\nonumber
<\sin \psi ImY^L_M> =  \frac{1}{2\pi} \int I(\Omega)\sin \psi ImY^L_M(\Omega d\Omega d\psi & = & \frac{1}{4}r^L_M\\ 
\nonumber
<ImY^L_M> = \frac{1}{2\pi} \int I(\Omega) ImY^L_M(\Omega d\Omega d\psi & = & \frac{1}{4}q^L_M
\end{eqnarray}
where $E_0=1$ and $E_M=2$ for $M\neq 0$. In terms of density matrix elements the parameters $t,p,r.q$ read~\cite{lutz78,sakrejda84}
\begin{eqnarray}
t^L_M & = & \sum \limits_J \sum \limits_{J'\lambda'} K^{LM}_{JJ'\lambda'}Re \bigl(R^0_u \bigr)^{J,J'}_{M+\lambda',\lambda'}\\
\nonumber
p^L_M & = & \sum \limits_J \sum \limits_{J'\lambda'} K^{LM}_{JJ'\lambda'}Re \bigl(R^0_y \bigr)^{J,J'}_{M+\lambda',\lambda'}\\
\nonumber
r^L_M & = & \sum \limits_J \sum \limits_{J'\lambda'} K^{LM}_{JJ'\lambda'}Im \bigl(R^0_x \bigr)^{J,J'}_{M+\lambda',\lambda'}\\
\nonumber
q^L_M & = & \sum \limits_J \sum \limits_{J'\lambda'} K^{LM}_{JJ'\lambda'}Im \bigl(R^0_z \bigr)^{J,J'}_{M+\lambda',\lambda'}
\end{eqnarray}
where
\begin{equation}
K^{LM}_{JJ'\lambda'}=(-1)^{\lambda'} \sqrt{\frac{(2J+1)(2J'+1)}{4\pi (2L+1)}}<JJ'00|L0><JJ'M+\lambda'-\lambda'|LM>
\end{equation}
General expressions for the density matrix elements $(R^0_k)^{JJ'}_{\lambda \lambda'},k=u,y,x,z$ in terms of the unnatural and natural exchange transversity amplitudes $U^J_{\lambda,\tau}$ and $N^J_{\lambda,\tau}$ are given in the Table I. Here $\tau=+\frac{1}{2},-\frac{1}{2}=up(u),down(d)$ is the target nucleon transversity. General expressions for the full set of density matrix elements $(R^j_k)^{JJ'}_{\lambda \lambda'},k=u,y,x,z$ including recoil nucleon polarization $j=1,2,3$ in terms of the unnatural and natural exchange transversity amplitudes are given in Ref.~\cite{svec13a,lutz78}.

\begin{table}
\caption{Density matrix elements $(R^0_k)^{JJ'}_{\lambda \lambda'},k=u,y,x,z$ expressed in terms of nucleon transversity amplitudes with definite $t$-channel naturality. The spin indices $JJ'$ which always go with helicities $\lambda \lambda'$ have been omitted in the amplitudes. The coefficients $\eta_\lambda=1$ for $\lambda=0$ and $\eta_\lambda=1/\sqrt{2}$ for $\lambda \neq 0$. Table from Ref.~\cite{svec13a}.}
\begin{tabular}{|l|r|}
\toprule
$(R^0_u)^{JJ'}_{\lambda \lambda'}$ & $\eta_\lambda \eta_{\lambda'} [U_{\lambda,u}U^*_{\lambda',u}+N_{\lambda,u}N^*_{\lambda',u}
+U_{\lambda,d}U^*_{\lambda',d}+N_{\lambda,d}N^*_{\lambda',d}]$\\
$(R^0_y)^{JJ'}_{\lambda \lambda'}$ & $\eta_\lambda \eta_{\lambda'} [U_{\lambda,u}U^*_{\lambda',u}+N_{\lambda,u}N^*_{\lambda',u}
-U_{\lambda,d}U^*_{\lambda',d}-N_{\lambda,d}N^*_{\lambda',d}]$\\
$(R^0_x)^{JJ'}_{\lambda \lambda'}$ & $-i\eta_\lambda \eta_{\lambda'} [U_{\lambda,u}N^*_{\lambda',d}+N_{\lambda,u}U^*_{\lambda',d}
-U_{\lambda,d}N^*_{\lambda',u}-N_{\lambda,d}U^*_{\lambda',u}]$\\
$(R^0_z)^{JJ'}_{\lambda \lambda'}$ & $\eta_\lambda \eta_{\lambda'} [U_{\lambda,u}N^*_{\lambda',d}+N_{\lambda,u}U^*_{\lambda',d}
+U_{\lambda,d}N^*_{\lambda',u}+N_{\lambda,d}U^*_{\lambda',u}]$\\

\botrule
\end{tabular}
\label{Table I.}
\end{table}

\section{The $S$-, $P$- and $D$-wave subsystem in $\pi^- p \to \pi^- \pi^+ n$.}

The $S$-,$P$- and $D$-wave subsystem is described by parameters $t,p,r,q$ for $L \leq 4$ and $M \leq 4$. The CERN measurements on transversely polarized target did not measure the parameters $q^L_M$. Expressions for $t,p,r$ in terms of the transversity amplitudes of definite naturality for $L\leq 4$ and $M \leq 2$ corresponding to $J\leq 2$ and $\lambda \leq 1$ were given by Lutz and Rybicki in Ref.~\cite{lutz78}. Expressions for $t,p,r$ for $L\leq 4$ and $M \leq 4$ corresponding to $J\leq 2$ and $\lambda \leq 2$ were given by Sakrejda in Ref.~\cite{sakrejda84}. 

In this work we focus on the parameters $t^L_M$ and $p^L_M$. These parameters organize themselves into two groups: $t^L_M+p^L_M$ are expressed in terms of bilinear terms $Re(A_uB^*_u)$ with transversity $up$, while  $t^L_M-p^L_M$ are expressed in terms of bilinear terms $Re(A_dB^*_d)$ with transversity $down$. We define the following convenient set of observables $a_{i,\tau}. i=1,15$
\begin{eqnarray}
a_{1,\tau}=\sqrt{\pi}(t^0_0 \pm p^0_0) & , & a_{2,\tau}=\sqrt{\pi}(t^2_0 \pm p^2_0)\sqrt{5}\\
\nonumber
a_{3,\tau}=\sqrt{\pi}(t^2_2 \pm p^2_2)\bigl(-\sqrt{\frac{5}{6}} \bigr) & , & a_{4,\tau}=\sqrt{\pi}(t^1_0 \pm p^1_0)\frac{1}{2}\\
\nonumber
a_{5,\tau}=\sqrt{\pi}(t^2_1 \pm p^2_1)\bigl(\frac{1}{2}\sqrt{\frac{5}{6}} \bigr) & , & a_{6,\tau}=\sqrt{\pi}(t^1_1 \pm p^1_1)\bigl 
(\frac{1}{2}\sqrt{\frac{1}{2}}\bigr )
\end{eqnarray}
\begin{eqnarray}
a_{7,\tau}=\sqrt{\pi}(t^3_0 \pm p^3_0)\bigl(\frac{1}{6}\sqrt{\frac{35}{3}} \bigr) & , & a_{8,\tau}=\sqrt{\pi}(t^3_1 \pm p^3_1)\bigl(\frac{1}{8}\sqrt{\frac{35}{3}}\bigr)\\
\nonumber
a_{9,\tau}=\sqrt{\pi}(t^3_2 \pm p^3_2)\bigl(\frac{1}{2}\sqrt{\frac{7}{6}} \bigr) & , & a_{10,\tau}=\sqrt{\pi}(t^4_0 \pm p^4_0)\frac{7}{2}\\
\nonumber
a_{11,\tau}=\sqrt{\pi}(t^4_1 \pm p^4_1)\bigl(\frac{7}{4}\sqrt{\frac{1}{35}} \bigr) & , & a_{12,\tau}=\sqrt{\pi}(t^4_2 \pm p^4_2)\bigl(\frac{7}{2}\sqrt{\frac{1}{10}}\bigr)
\end{eqnarray}
\begin{eqnarray}
a_{13,\tau}=\sqrt{\pi}(t^3_3 \pm p^3_3)\bigl(\frac{\sqrt{7}}{3} \bigr) & , & a_{14,\tau}=\sqrt{\pi}(t^4_3 \pm p^4_3)\bigl( \sqrt{\frac{7}{5}} \bigr)\\
\nonumber
a_{15,\tau}=\sqrt{\pi}(t^4_4 \pm p^4_4)\bigl( \sqrt{\frac{14}{5}} \bigr)
\end{eqnarray}
In (3.1)-(3.3) $\tau=u$ for the $+$ sign and $\tau=d$ for the $-$ sign. Next we express the observables $a_{i,\tau}$ in terms of $S$-, $P$- and $D$-wave amplitudes defined as follows
\begin{eqnarray}
\begin{array} {lll}
U^0_{0,\tau}=S_\tau &  &  \\
U^1_{0,\tau}=L_\tau & U^1_{1,\tau}=U_\tau & N^1_{1,\tau}=N_\tau\\
U^2_{0,\tau}=D^0_\tau & U^2_{1,\tau}=D^U_\tau & U^2_{2,\tau}=D^{2U}_\tau\\
N^2_{1,\tau}=D^N_\tau & N^2_{2,\tau}=D^{2N}_\tau &   \\
\end{array}
\end{eqnarray}
For the purposes of our analysis we shall split the observables $a_{i,\tau}$ into three parts
\begin{equation}
a_{i,\tau}=c_{i,\tau}+d_{i,\tau}+e_{i,\tau}
\end{equation}
where $c_{i,\tau}$ involve only $S$- and $P$-wave amplitudes, $d_{i,\tau}$ involve terms with $D$-wave amplitudes with only helicity $\lambda \leq 1$, and $e_{i,\tau}$ involve terms with $D$-wave amplitudes with $\lambda=2$ (rank 2 amplitudes). The expressions for the $D$-wave terms $d_{i,\tau}$ and $e_{i,\tau}$ in terms of the transversity amplitudes are given in the Table II. The expressions for $c_{i,\tau}$ read as follows
\begin{eqnarray}
c_{1,\tau} & = & |S_\tau|^2+|L_\tau|^2 +|U_\tau|^2+|N_\tau|^2 \\
\nonumber
c_{2,\tau} & = & 2|L_\tau|^2-|U_\tau|^2-|N_\tau|^2 \\
\nonumber
c_{3,\tau} & = & |N_\tau|^2-|U_\tau|^2 \\ 
\nonumber
c_{4,\tau} & = & |L_\tau||S_\tau|\cos \Phi(L_\tau S^*_\tau) \\
\nonumber
c_{5,\tau} & = & |L_\tau||U_\tau|\cos \Phi(L_\tau U^*_\tau) \\
\nonumber
c_{6,\tau} & = & |U_\tau||S_\tau|\cos \Phi(U_\tau S^*_\tau) 
\end{eqnarray}
where the cosines of relative phases
\begin{equation}
\cos \Phi(A_\tau B^*_\tau)=\cos(\Phi(A_\tau)-\Phi(B_\tau))
\end{equation}
All $c_{i,\tau}=0$ for $i=7,15$.

\begin{table}
\caption{$D$-wave contributions $d_{i,\tau}$ and $e_{i,\tau}$ to the observables $a_{i,\tau}$ corresponding to $D$-wave transversity amplitudes with helicities $\lambda \leq 1$ and $\lambda \leq 2$, respectively. The transversity index $\tau$ is omitted for the sake of brevity and the bilinear terms $AB^* \equiv Re(AB^*)$. Table from Ref.~\cite{lutz78,sakrejda84}.}
\begin{tabular}{|c|c|c|}
\toprule
$a_{i,\tau}$ & $d_{i,\tau}$ & $e_{i,\tau}$ \\
\colrule
$a_1$ & $|D^0|^2+|D^U|^2+|D^N|^2$ & $|D^{2U}|^2+|D^{2N}|^2$ \\
$a_2$ & $2\sqrt{5}D^0S^*+\frac{5}{7}(2|D^0|^2+|D^U|^2+|D^N|^2)$ & $-\frac{10}{7}(|D^{2U}|^2+|D^{2N}|^2)$ \\
$a_3$ & $\frac{5}{7}(|D^N|^2-|D^U|^2)$ & $-2\sqrt{\frac{5}{3}} SD^{2U*}+\frac{20}{7}\sqrt{\frac{1}{3}}D^0 D^{2U*}$ \\
$a_4$ & $\sqrt{\frac{4}{5}}D^0L^*+\sqrt{\frac{3}{5}}(D^UU^*+D^NN^*)$ & 0 \\
$a_5$ & $\sqrt{\frac{5}{3}}D^US^*+\frac{5}{7}\sqrt{\frac{1}{3}}D^UD^{0*}$ & 
$\frac{5}{7}(D^UD^{2U*}+D^ND^{2N*})$ \\
$a_6$ & $\sqrt{\frac{3}{5}}D^UL^*-\sqrt{\frac{1}{5}}D^0U^*$ & $\sqrt{\frac{3}{5}}(UD^{2U*}+ND^{2N*})$ \\
\colrule
$a_7$ & $D^0L^*-\sqrt{\frac{1}{3}}(D^UU^*+D^NN^*)$ & 0 \\
$a_8$ & $D^UL^*+\sqrt{\frac{3}{4}}D^0U^*$ & $-\frac{1}{4}(UD^{2U*}+ND^{2N*})$ \\
$a_9$ & $D^UU^*-D^NN^*$ & $LD^{2U*}$ \\
$a_{10}$ & $3|D^0|^2-2(|D^U|^2+|D^N|^2)$ & $\frac{1}{2}(|D^{2U}|^2+|D^{2N}|^2)$ \\
$a_{11}$ & $D^UD^{0*}$ & $-\frac{1}{2}\sqrt{\frac{1}{7}}(D^UD^{2U*}+D^ND^{2N*})$ \\
$a_{12}$ & $|D^U|^2-|D^N|^2$ & $\sqrt{3}D^0D^{2U*}$ \\
\colrule
$a_{13}$ & 0 & $UD^{2U*}-ND^{2N*}$ \\
$a_{14}$ & 0 & $D^UD^{2U*}-D^ND^{2N*}$ \\
$a_{15}$ & 0 & $|D^{2U}|^2-|D^{2N}|^2$ \\

\botrule
\end{tabular}
\label{Table II.}
\end{table}

Finally we present expressions for the parameters $r^L_M$. We define new observables
\begin{eqnarray}
r_1 & = & -\frac{1}{4}\sqrt{4\pi} r^1_1= (R^0_x)^{10}_{1s}+r_1(D)\\
\nonumber
r_2 & = & -\frac{1}{2\sqrt{2}}\sqrt{\frac{5}{6}}\sqrt{4\pi} r^2_1= (R^0_x)^{11}_{10}+r_2(D)\\
\nonumber
r_3 & = & +\frac{1}{2}\sqrt{\frac{5}{6}}\sqrt{4\pi} r^2_2= (R^0_x)^{11}_{1-1}+r_3(D)
\end{eqnarray}
where $R^0_x$ are the density matrix elements~\cite{svec13a,lutz78,sakrejda84}
\begin{eqnarray}
(R^0_x)^{10}_{1s} & = & -\frac{1}{\sqrt{2}} Re\bigl(N_dS_u^*-N_uS_d^*\bigr)\\
\nonumber
(R^0_x)^{11}_{10} & = & -\frac{1}{\sqrt{2}} Re\bigl(N_dL_u^*-N_uL_d^*\bigr)\\
\nonumber
(R^0_x)^{11}_{1-1} & = & + Re\bigl(N_dU_u^*-N_uU_d^*)
\end{eqnarray}
The $D$-wave contributions $r_k(D),k=1,3$ are given by~\cite{lutz78,sakrejda84}
\begin{eqnarray}
r_1(D) & = & -\frac{1}{\sqrt{10}} Re\bigl(N_dD^{0*}_u-N_uD^{0*}_d\bigr)+\sqrt{\frac{3}{10}} Re\bigl(D^N_dL^*_u-D^N_uL^*_d\bigr)\\
\nonumber
   &    & +\frac{1}{2}\sqrt{\frac{6}{5}} Re\bigl(N_dD^{2U*}_u-N_uD^{2U*}_d\bigr)-\frac{1}{2}\sqrt{\frac{6}{5}} Re\bigl(D^{2N}_dU^*_u-D^{2N}_uU^*_d\bigr)\\
\nonumber
r_2(D) & = & -\sqrt{\frac{5}{6}} Re\bigl(D^N_dS^*_u-D^N_uS^*_d\bigr)\\
\nonumber
r_3(D) & = & \sqrt{\frac{3}{5}} Re\bigl( D^{2N}_dS^*_u-D^{2N}_uS^*_d\bigr)
\end{eqnarray}

\section{Observed bilinear terms and the Kraus amplitudes.}

The expressions for the observed parameters $t,p,r$ in terms of the transversity amplitudes presented in the previous Section were derived using a unitary evolution law (1.1). To be physically meaningfull the observed density matrix $\rho_f(O)$ must have the same bilinear structure as the final state density matrix $\rho_f(S)$ produced by the $S$-matrix dynamics. The observed bilinear terms are then related to the bilinear terms of Kraus amplitudes by (1.4)~\cite{svec13b} which in turn are related to the $S$-matrix bilinear terms via the the unitary transforms involving the matrix elements of the Kraus operators. In general, Kraus transversity amplitudes are related to $S$-matrix transversity amplitudes by a unitary transform~\cite{svec13b}
\begin{equation}
A^{J}_{\lambda,\tau}(\ell)= \sum \limits_{K=J-1,J,J+1}<J\lambda,+|V_\ell|K\lambda,+>A^{K}_{\lambda,\tau}(S)
\end{equation}
where $+$ stands for the recoil nucleon helicity $+\frac{1}{2}$. In $\pi^- p \to \pi^- \pi^+ n$ and $\pi^+ n \to \pi^+ \pi^- p$ below 1400 MeV the relation (4.1) reduces to relations (1.5)-(1.7). In $\pi^- p \to \pi^0 \pi^0 n$ and $\pi^+ p \to \pi^+ \pi^+ n$ the relation (4.1) reduces to (1.5).

For two decoherence free amplitudes $A$ and $B$ that mix spins the relation (1.4) reduces to
\begin{equation}
|A(O)||B(O)|\cos \Phi(A(O)B(O)^*)=|A||B|\cos (\phi(A)-\phi(B))
\end{equation}
so that the observed amplitudes are equal to the spin mixing Kraus amplitudes. This expression also applies when only one of the amplitudes is mixing spins. For two decoherence free amplitudes that do not mix spins we have
\begin{equation}
|A(O)||B(O)|\cos \Phi(A(O)B(O)^*)=Re(A^0B^{0*})\cos (\phi(\alpha)-\phi(\beta))
\end{equation}
where $\phi(\alpha)=\alpha+\alpha^0$, $\phi(\beta)=\beta+\beta^0$ are the full dephasing phases of $A$ and $B$, and where $A^0,\alpha^0$ and $B^0,\beta^0$ are the corresponding $S$-matrix amplitudes and their phases. 

For two decohering amplitudes $A$ and $B$ the relation (1.4) reads
\begin{equation}
|A(O)||B(O)|\cos \Phi(A(O)B(O)^*)=Re(A^0B^{0*})\sum \limits_{\ell=1}^M 
p_{\ell \ell}\cos(\phi(\alpha(\ell))-\phi(\beta(\ell)))
\end{equation}
The phases $\phi(\alpha(\ell))$ and $\phi(\beta(\ell))$ are the $\ell$-dependent full phases of Kraus amplitudes $A(\ell)$ and $B(\ell)$. A bilinear term of a decohering amplitude $A$ and a decoherence free amplitude $B$ has the form
\begin{equation}
|A(O)||B(O)|\cos \Phi(A(O)B(O)^*)=|A^0||B|\sum \limits_{\ell=1}^M p_{\ell \ell}
\cos (\phi(A(\ell))-\phi(B))
\end{equation}
where $\phi(A(\ell))$ and $\phi(B)$ are the full phases of $A(\ell)$ and $B$ including the phases of the $S$-matrix amplitudes.

Kraus amplitudes are complex valued functions that always satisfy phase relation
\begin{equation}
(\phi(A)-\phi(B))-(\phi(A)-\phi(C))-(\phi(C)-\phi(B))=0
\end{equation}
for any three amplitudes $A$, $B$ and $C$. With $\Phi(AB^*)=\phi(A)-\phi(B)$, the equivalent condition is the cosine condition
\begin{equation}
\cos^2\Phi(AB^*)+\cos^2\Phi(AC^*)+\cos^2\Phi(CB^*)-
2\cos\Phi(AB^*)\cos\Phi(AC^*)\cos\Phi(CB^*)=1
\end{equation}
The observed decoherence free amplitudes are complex valued functions and thus satisfy the phase and cosine conditions. The observed decohering amplitudes are no longer complex valued functions with $\cos \Phi(A(O)B(O)^*)$ having a simple meaning of a correlation factor. These correlation factors violate the cosine conditions with r.h.s. of (4.7) equal to $1+G$ where $G$ is called cosine gap. Similarly, the "phases" $\Phi(AB^*)$, $\Phi(AC^*)$ and $\Phi(CB^*)$ violate the phase conditions with r.h.s. of (4.6) equal to $\Delta$ where $\Delta$ is called a phase gap. 

Kraus aplitudes must satisfy exact cosine conditions. This allows us to define Kraus amplitudes as the analytical solutions arising from imposing exact cosine condition(s) either on a suitable subset of the measured observables or on solutions for their phases arising from relations between their bilinear terms (Sections V. and VI.).

\section{Amplitude analysis of the $S$- and $P$-wave subsystem below 1080 MeV.}

\subsection{The inversion of the spin mixing mechanism}

Expressed in terms of the spin mixing mechanism (1.8) and Kraus amplitude (1.9) the ten $S$-and $P$-wave bilinear terms of each transversity read
\begin{eqnarray}
A_\tau = |L_\tau|^2+|S_\tau|^2 & = & |L^0_\tau|^2+|S^0_\tau|^2\\
\nonumber
B_\tau = |L_\tau|^2-|S_\tau|^2 & = & -\cos 2\theta (|L^0_\tau|^2-|S^0_\tau|^2
+2\sin 2\theta \cos \phi |L^0_\tau||S^0_\tau|\\
\nonumber
|U_\tau|^2 & = & |U^0_\tau|^2\\
\nonumber
|N_\tau|^2 & = & |N^0_\tau|^2\\
c_{4,\tau} = Re(L_\tau S_\tau^*) & = & 0.5 \sin 2\theta (|L^0_\tau|^2-
|S^0_\tau|^2+\cos 2\theta \cos \phi |L^0_\tau||S^0_\tau|\\
\nonumber
c_{5,\tau} =  Re(L_\tau U_\tau^*) & = & -\cos \theta \cos \phi |S^0_\tau||U^0_\tau|
-\sin \theta |L^0_\tau||U^0_\tau|\\
\nonumber
c_{6,\tau} =  Re(U_\tau S_\tau^*) & = & +\sin \theta \cos \phi |S^0_\tau||U^0_\tau|
-\cos \theta |L^0_\tau||U^0_\tau|\\
g_{4,\tau} = Im(L_\tau S_\tau^*) & = & -\sin \phi |L^0_\tau||S^0_\tau|\\
\nonumber
g_{5,\tau} =  Im(L_\tau U_\tau^*) & = & +\sin \phi \cos \theta |L^0_\tau||U^0_\tau|\\
\nonumber
g_{6,\tau} =  Im(U_\tau S_\tau^*) & = & +\sin \phi \sin \theta |S^0_\tau||U^0_\tau|
\end{eqnarray}
In the derivation of the equations (5.1)-(5.3) we have used a set of self-consistent relative phases of the $S$-matrix amplitudes given in the Table II. of Ref.~\cite{svec13a}. After some algebra the equations (5.1) and (5.2) can be solved for 
$\sin \theta$
\begin{equation}
\sin \theta = \frac{||S_\tau|^2 R_\tau +c_{4,\tau}|}
{\sqrt{\bigl(|S_\tau|^2 R_\tau +c_{4,\tau} \bigr)^2 +\bigl(|L_\tau|^2  + c_{4,\tau} R_\tau \bigr)^2}}
\end{equation}
where the ratio 
\begin{equation}
R_\tau = -\frac{c_{5,\tau}}{c_{6,\tau}}
\end{equation}
The solutions for the $S$-matrix moduli $|S^0_\tau|^2$ and $|L^0_\tau|^2$ read
\begin{eqnarray}
|L^0_\tau|^2 & = & |L_\tau|^2 \sin^2 \theta + |S_\tau|^2 \cos^2 \theta +
|L_\tau||S_\tau|\cos \Phi_\tau(LS) \sin 2\theta\\
\nonumber
|S^0_\tau|^2 & = & |L_\tau|^2 \cos^2 \theta + |S_\tau|^2 \sin^2 \theta -
|L_\tau||S_\tau|\cos \Phi_\tau(LS) \sin 2\theta
\end{eqnarray}
Note that these solutions do not depend on the parameter $\phi$. The solution for $\phi$ is given by
\begin{eqnarray}
\cos \phi & = & \frac{B_\tau \sin 2\theta + 2 c_{4,\tau} \cos 2\theta}
{|L^0_\tau||S^0_\tau|}\\
\nonumber
\sin \phi & = & -\frac{g_{4,\tau}}{|L^0_\tau||S^0_\tau|}
\end{eqnarray}
The independence of the r.h.s. of the equation (5.4) on $\tau$ requires that $R_d=R_u=R$ and the scaling relations for the spin mixing amplitudes
\begin{eqnarray}
|S_u| & = & K |S_d|\\
\nonumber
|L_u| & = & K |L_d|
\end{eqnarray}
with $\tau$ independent phase $\Phi(L_\tau S_\tau^*)$. Together with (5.8) the condition $R_d=R_u=R$ implies $\tau$ independent phases $\Phi(L_\tau U_\tau^*)$ and $\Phi(U_\tau S_\tau^*)$. The relations (5.6) imply the same scaling relations (5.8) also for the $S$-matrix amplitudes $|S^0_\tau|$ and 
$|L^0_\tau|$.

\subsection{Observed and $S$-matrix transversity amplitudes for transversity $\tau=d$}

We perform a new amplitude analysis of the CERN measurements of the observables $t^L_M,p^L_M,r^L_M,L\leq 2,M\leq 2$ in $\pi^- p \to \pi^- \pi^+ n$ on polarized target at 17.2 GeV/c at $|t| \leq 0.20$ (GeV/c)$^2$ using spin mixing mechanism. In our Monte Carlo amplitude analysis each sampling of the data error volume defines two groups of parameters $a_{k,u}$ and $a_{k,d}$, $k=1,6$. Previous amplitude analyses of this data~\cite{kaminski97,svec13b} established that the $\tau=d$ $S$- and $P$-wave transversity amplitudes are about three times larger than the corresponding $\tau=u$ amplitudes. This allows us to neglect the $D$-wave contributions to $\tau=d$ observables and to assume that the amplitudes are Kraus amplitudes.

With $a_{k,d}=c_{k,d}$ we can use (3.6) to solve for the moduli $|S_d|^2,|U_d|^2,|N_d|^2$ in terms of $|L_d|^2$
\begin{eqnarray}
|S_d|^2 & = & a_{1,d}+a_{2,d}-3|L_d|^2\\
\nonumber
|U_d|^2 & = & |L_d|^2-\frac{1}{2}(a_{2,d}+a_{3,d})\\
\nonumber
|N_d|^2 & = & |L_d|^2-\frac{1}{2}(a_{2,d}-a_{3,d})
\end{eqnarray}
and for the cosines of the relative phases in terms of the moduli. From the cosine condition we obtain a cubic equation for $|L_d|^2$ that has two physical solutions of the form~\cite{svec13b}
\begin{eqnarray}
|L_d(1)|^2 & = & |L_{d,0}|^2+\Delta_d\\
\nonumber
|L_d(2)|^2 & = & |L_{d,0}|^2-\Delta_d
\end{eqnarray}
which lead to two solutions for the moduli and two solutions for the cosines. For each solution for the moduli there are two solutions for the relative phases differing in the sign due to the ambiguity $\cos \Phi = \cos (\pm \Phi)$.
We work with the solutions $\Phi(L_d S_d^*) \geq 0$.

The two solutions for the moduli and phases for the $\tau=d$ observed Kraus transversity amplitudes define two sets of the bilinear terms (5.1)-(5.3) for $\tau=d$ which in turn lead to two solutions for the spin mixing parameters $\theta$ and $\phi$ and $S$-matrix transversity amplitudes $|S^0_d|^2$, $|L^0_d|^2$, $|U^0_d|^2$ and $|N^0_d|^2$. 

\subsection{Observed and $S$-matrix transversity amplitudes for transversity $\tau=u$}

In our previous study~\cite{svec12d} of the effect of the $D$-wave amplitudes on the $S$- and $P$-wave amplitude analysis of the CERN data we assumed that $D$-wave amplitudes contribute only to the observables $a_{k,\tau}, k=1,3$. For small $D$-wave amplitudes we found only a very small effect. This result allows us to neglect $D$-wave amplitudes in these observables so that we shall assume 
$a_{k,u}=c_{k,u}, k=1,3$. This relation can only be approximate for $a_{1,u}$ since we are going to assume that the principal effect of the $D$-waves involves the observables $a_{k,u},k=4,6$. This means we can no longer use the cosine condition involving these observables to calculate the two solutions for $|L_u|^2$.

\begin{figure} [htp]
\includegraphics[width=12cm,height=10.5cm]{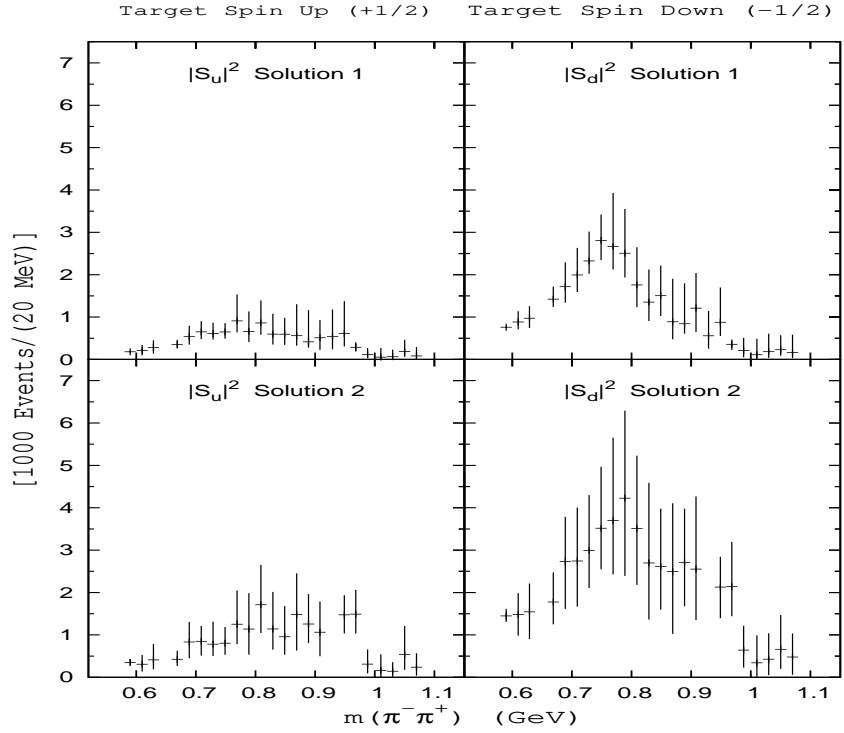}
\caption{Solutions 1 and 2 for the observed $S$-wave spin mixing Kraus amplitudes $|S_u|^2$ and $|S_d|^2$.}
\label{Figure 1}
\end{figure}

\begin{figure} [hp]
\includegraphics[width=12cm,height=10.5cm]{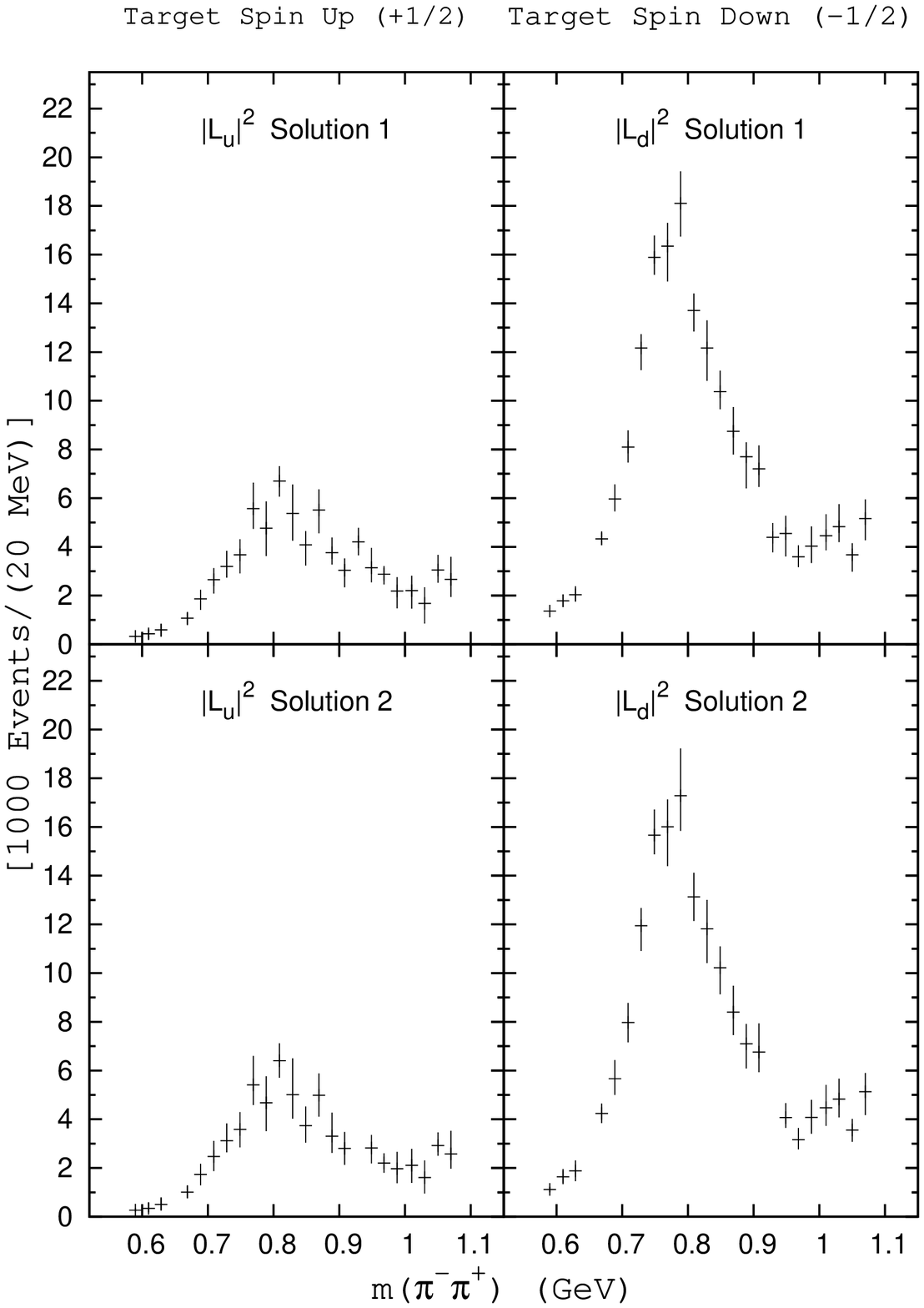}
\caption{Solutions 1 and 2 for the observed $P$-wave spin mixing Kraus amplitudes $|L_u|^2$ and $|L_d|^2$.}
\label{Figure 2}
\end{figure}

\begin{figure} [htp]
\includegraphics[width=12cm,height=10.5cm]{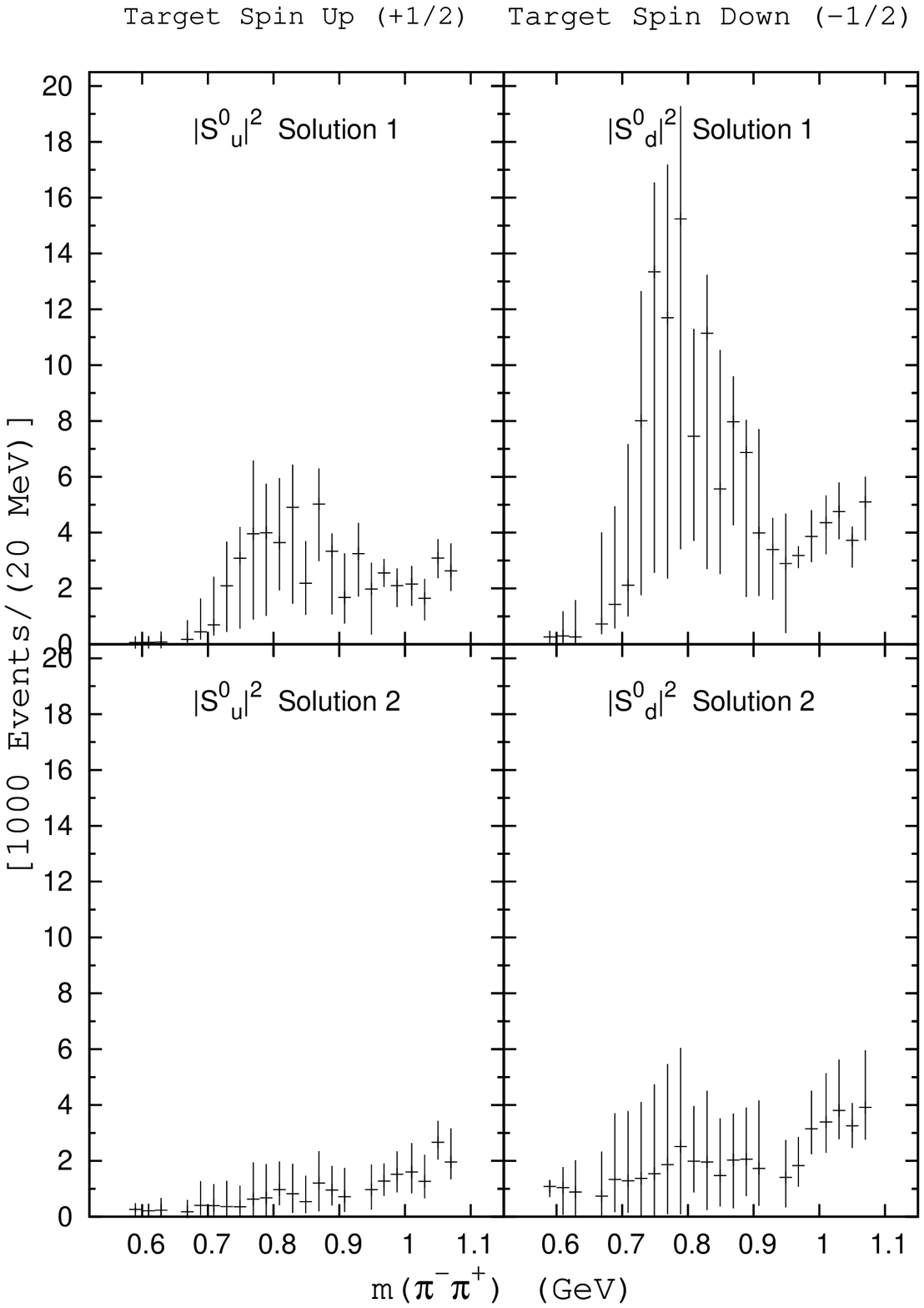}
\caption{Solutions 1 and 2 for the $S$-wave $S$-matrix transversity amplitudes $|S^0_u|^2$ and $|S^0_d|^2$.}
\label{Figure 3}
\end{figure}

\begin{figure} [hp]
\includegraphics[width=12cm,height=10.5cm]{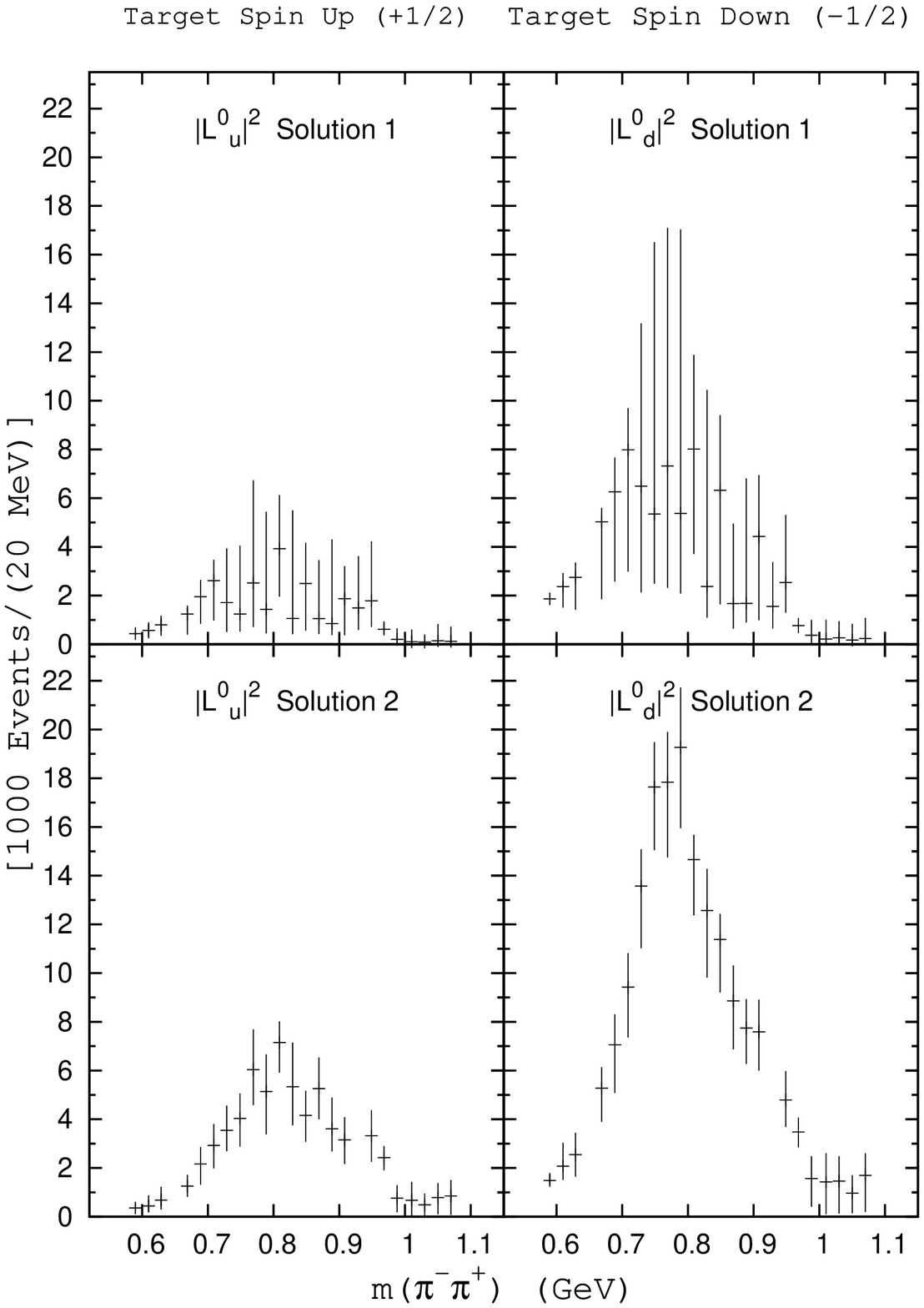}
\caption{Solutions 1 and 2 for the $P$-wave $S$-matrix transversity amplitudes $|L^0_u|^2$ and $|L^0_d|^2$.}
\label{Figure 4}
\end{figure}

\begin{figure} [htp]
\includegraphics[width=12cm,height=10.5cm]{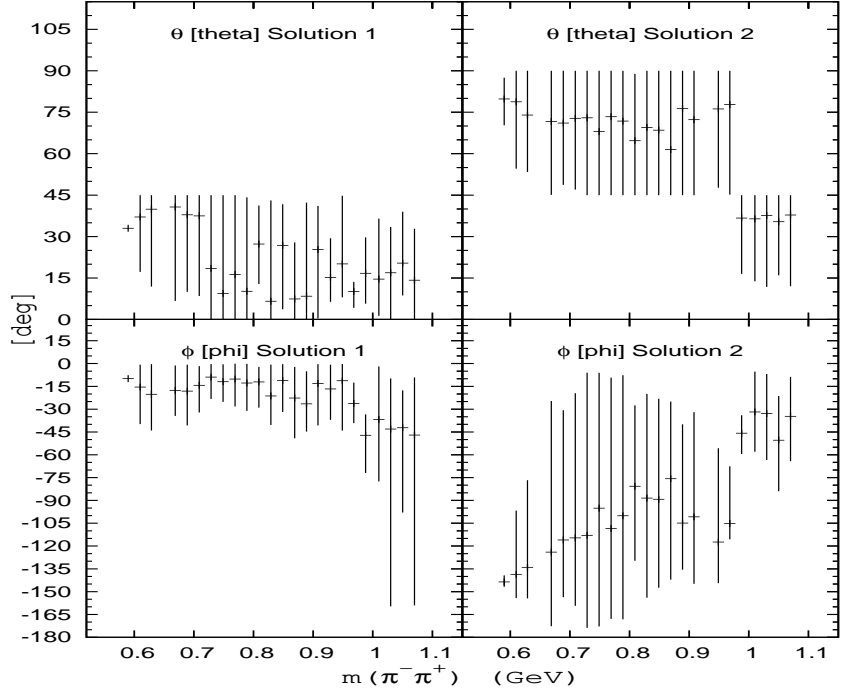}
\caption{Solutions 1 and 2 for the spin mixing parameters $\theta$ and $\phi$.}
\label{Figure 5}
\end{figure}

\begin{figure} [hp]
\includegraphics[width=12cm,height=10.5cm]{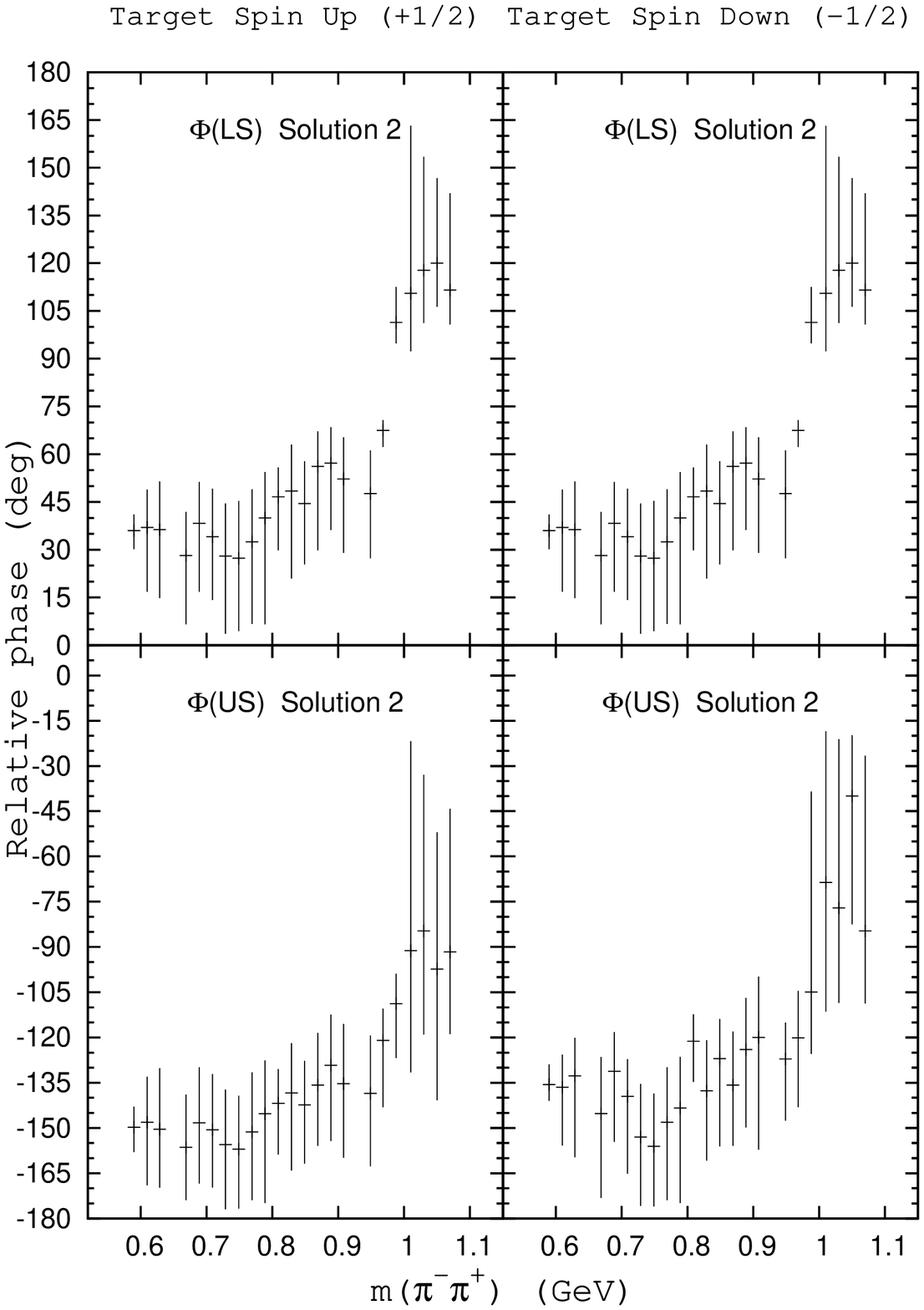}
\caption{Solution 2 for the relative phases $\Phi(LS)=\phi(L)-\phi(S)$ and $\Phi(US)=\phi(U)-\phi(S)$.}
\label{Figure 6}
\end{figure}

Given our assumptions we still have relations for the moduli of Kraus transversity amplitudes
\begin{eqnarray}
|S_u|^2 & = & a_{1,u}+a_{2,u}-3|L_u|^2\\
\nonumber
|U_u|^2 & = & |L_u|^2-\frac{1}{2}(a_{2,u}+a_{3,u})\\
\nonumber
|N_u|^2 & = & |L_u|^2-\frac{1}{2}(a_{2,u}-a_{3,u})
\end{eqnarray}
For $|S_u|^2$ and $|L_u|^2$ we use the scaling relations (5.8). From the first equation in (5.11) we find
\begin{equation}
K^2=\frac{a_{1,u}+a_{2,u}}{a_{1,d}+a_{2,d}}
\end{equation}
Thus from the two solutions for $|L_d|^2$ we obtain also two corresponding solutions for $|L_u|^2,|S_u|^2$ and $|U_u|^2=|U^0_u|^2,|N_u|^2=|N^0_u|^2$. Using the scaling relations $|S^0_u|^2=K^2|S^0_d|^2$ and $|L^0_u|^2=K^2|L^0_d|^2$ we find the moduli of all $S$-matrix transversity amplitudes. From (5.2) we find the $S$- and $P$-wave interference terms $c_{k,u},k=4,6$. It is these terms that satisfy the cosine condition. They differ from the input observables $a_{k,u},k=4,6$
\begin{equation}
\Delta_{k,u}=a_{k,u}-c_{k,u}=e_{k,u}+d_{k,u}
\end{equation}
where the observables $\Delta_{k,u},k=4,6$ describe the $D$-wave contributions $e_{k,u}+d_{k,u}$.

\subsection{Results for the spin mixing and $S$-matrix transversity ampitudes}

In Section VI. we use the new observables $\Delta_{k,u},k=4,6$ together with the assumptions $\Delta_{k,u}=0,k=2,3$ and $a_{k,u}=0$ or $a_{k,u} \sim 0$ for $k=7-15$ to perform a model dependent $D$-wave amplitude analysis. This analysis imposes additional constraints on the two solutions for the $S$- and $P$-wave spin mixing amplitudes. We have performed the $S$- and $P$-wave amplitude analysis with and without the $D$-wave analysis. The results are nearly identical. In this Paper we report the results with the $D$-wave analysis.

The two solutions for the observed spin mixing Kraus aplitudes $|S_\tau|^2$ and $|L_\tau|^2$ are shown in Figures 1 and 2. In agreement with previous analyses~\cite{svec12a}, the $S$-wave amplitude $|S_d|^2$ shows a clear $\rho^0(770)$ peak in both solutions while the $P$-wave amplitude 
$|L_d|^2$ shows a dip near $f_0(980)$ mass. The $S$-matrix amplitudes $|S^0_\tau|^2$ and $|L^0_\tau|^2$ shown in Figures 3 and 4 are dramatically different. 

In the Solution 1 the amplitude $|S^0_d(1)|^2$ is dominant as it resonates at the $\rho^0(770)$ mass having a character of the amplitude $|L_d(1)|^2$. The amplitudes $|L^0_u(1)|^2$ and  $|L^0_d(1)|^2$ show no clear resonant behavour but somewhat random structures instead. In contrast, in the Solution 2 the amplitude $|S^0_d(2)|^2$ is small, shows no evidence of the $\rho^0(770)$ mixing, and is rising above the $K \bar{K}$ threshold consistent with $f_0(980)$ resonance. Both amplitudes $|L^0_u(2)|^2$ and $|L^0_d(2)|^2$ show  clear resonant behavour at $\rho^0(770)$ mass. 

We conclude that the Solution 1 is excluded by the SMM while the Solution 2 is favoured. There is thus a single solution for the $S$- and $P$-wave Kraus amplitudes which implies that the $S$- and $P$-wave Kraus amplitudes form a decoherence free subsystem.

Recall that there is no spin mixing for $\theta=\pi/2$ which implies $|L_\tau|^2=|L^0_\tau|^2$ and $|S_\tau|^2=|S^0_\tau|^2$. A complete spin mixing occurs for $\theta=0$ when $|L_\tau|^2=|S^0_\tau|^2$ and $|S_\tau|^2=|L^0_\tau|^2$. We found that only a very few percent of the Monte Carlo physical Solutions 1 have $\theta > \pi/4$ and only a small percentage of the Monte Carlo physical Solutions 2 below 980 MeV have $\theta < \pi/4$. These minority values of $\theta$ were cut out. Above 980 MeV Monte Carlo Solutions 2 with $\theta > \pi/4$ were cut out to ensure that $|L^0_d|^2$ decreases and $|S^0_d|^2$ increases above the $K \bar{K}$ threshold.

Figure 5 shows the measured spin mixing parameters $\theta $ and $\phi$. In Solution 1 $\theta$ is somewhat random with the averaged values below $30^\circ$ indicating strong spin mixing while $\phi$ is nearly constant at $\phi \approx -15^\circ$. In Solution 2 the average values of $\theta$ are nearly constant at $\theta \approx 75^\circ$ below 980 MeV and at $\theta \approx 30^\circ$ above 980 MeV. This indicates stronger spin mixing in the $f_0(980)$ mass region. In the Solution 2 the average values of $\phi$ are increasing with dipion mass from $\sim -150^\circ$ to $\sim -75^\circ$.

Figure 6 shows the dependence of the relative phases $\Phi(L_\tau S_\tau^*)$ and $\Phi(U_\tau S_\tau^*)$ on the transversity $\tau$. Recall that for $\tau=d$ these phases are calculated from the random input data on observables $a_{k,d},k=4,6$ while for $\tau=u$ they are calculated from the bilinear terms $c_{k,u},k=4,6$ predicted by the SMM. Since SMM sets $a_{4,d}=c_{4,d}=c_{4,u}/{K^2}$ we obtain identity $\Phi(L_d S_d^*)=\Phi(L_u S_u^*)$. SMM does not impose similar relation for the observables $a_{5,d}$ and $a_{6,d}$ and only indirectly constrains these otherwise random observables. Yet the relative phases $\Phi(U_d S_d^*)$ and $\Phi(U_u S_u^*)$ are nearly equal as seen in the Figure 6. The same holds true for relative phases $\Phi(L_d U_d^*)$ and $\Phi(L_u U_u^*)$ (not shown). These results demonstrate the self-consistency of our new amplitude analysis based on SMM. 

The observed differences between the relative phases of opposite transversity have an important physical interptretation. In the calulations of bilinear terms $c_{k,u},k=4,6$ we have assumed that the amplitude $U_u$ does not mix with the $D$-wave amplitude $D^U_u$ and therefore it has the same phase $2\phi$ like the amplitude $U_d$.  Since the amplitude$D^U_u$ is present in the data this is an approximation. This amplitude $U_u$ should be replaced by a spin mixing amplitude $U'_u$ with a different phase $2\phi'$
\begin{eqnarray}
U_u=e^{i2\phi}U^0_u & \to & U'_u=e^{i2\phi'}U^{0,eff}_u
\end{eqnarray}
where $\phi'=\phi+\epsilon$ and $U^{0,eff}_u$ is an effective amplitude with the same phase as $U^0_u$. Then for suitable $\epsilon$ the relative phases from the new bilinear terms $c'_{k,u},k=5,6$ can be equated with the relative phases from $a_{k,d},k=5,6$ to ensure that $R_d=R_u$ in (5.4). In our analysis this condition is violated by the mixing in the amplitude $U_u$ with $R_d$ on average larger by $\sim 5\%$ than $R_u$. The small difference in the observed relative phases thus arises from the spin mixing in the amplitude $U_u$ which predicts $\rho^0(770)$ spin mixing in the amplitude $D^U_u$. This prediction is confirmed by our amplitude analysis of the $D$-wave subsystem (Figure 7).

The amplitude analysis does not make use of the paramaters $r^L_M$ as an input directly. Instead it predicts the parameters $r^1_1(th)$, $r^2_1(th)$ and $r^2_2(th)$ assuming no $D$-wave contributions, and imposes a constraint $\chi^2 \leq 3$ on the $\chi^2$ value of each $r^L_M$ for each Monte Carlo solution for the amplitudes. The results for the bin averaged values of such $\chi^2$ for each $r^L_M$ of the Solutions 1 and 2 with or without $D$-wave analysis are in all cases in the range 0.9 - 1.1. These low values demonstrate excellent predictions for the parameters $r^L_M$ and justify a posteriori the neglect of the $D$-wave contributions. The analysis is further constrained by the requirement that $|a_{k,u}-c_{k,u}| \leq 3 \sigma(a_{k,u})$ where $\sigma(a_{k,u})$ is the experimental error on $a_{k,u}$, $k=4,6$.

Finally we note that our analysis assumes the positive solution for $\Phi(LS^*)\geq 0$. The amplitudes for the negative solution are simply complex conjugate of the amplitudes with the positive phase. The two solutions differ experimentally only in the sign of the unmeasured observables $Im \rho^0_z$.
Our analysis predicts $Im (\rho^0_z)_{s1} \sim 0$, $Im (\rho^0_z)_{01}\sim 0$ and $Im (\rho^0_z)_{1-1}=0$ rendering the two phase solutions experimentally indistinguishable.

\section{Amplitude analysis of the $D$-wave subsystem below 1080 MeV.}

\subsection{Equations for the $\tau=u$ $D$-wave subsystem}

Identifying the observables $\Delta_{k,u}=a_{k,u}-c_{k,u},k=4,6$ with the $D$-wave contributions $\Delta_{k,u}=d_{k,u}+e_{k,u},k=4,6$ to $a_{k,u},k=4,6$ prompts us to similarly define $\Delta_{k,u}$ for $k=1,3$  with $\Delta_{2,u}=\Delta_{3,u}=0$. Together with the observables $a_{k,u},k=7,15$ the observables $\Delta_{k,u},k=2,6$ form a system of equations that separates from the $S$- and $P$-wave subsystem. The expressions for these observables in terms of amplitudes are given in the Table II. In this Section we shall omit the subscript $u$ for the sake of brevity of the notation.

We shall assume that the first term in $\Delta_2$  vanishes $2\sqrt{5}Re(D^0S^*)=0$. Then the dual pairs of observables $(\Delta_2,a_{10})$, $(\Delta_3,a_{12})$, $(\Delta_4,a_7)$, $(\Delta_5,a_{11})$, $(\Delta_6,a_8)$ involve bilinear terms that do not occur anywhere else and can be solved. The five groups of equations for the $\tau=u$ $D$-wave subsystem required for our amplitude analysisis then takes the form
\begin{eqnarray}
ReD^0S^* & = & 0\\
\nonumber
ReD^0L^* & = & \frac{1}{\sqrt{5}}\Delta_4 + \frac{3}{4} a_7\\
ReD^US^* & = & \sqrt{\frac{3}{5}}\Delta_5+\frac{\sqrt{5}}{7}\bigl(2\sqrt{21}a_{11} - (2\sqrt{21}+1)ReD^UD^{0*} \bigr)\\
\nonumber
ReD^UL^* & = & \frac{1}{15}\Delta_6-\frac{1}{3}ReD^0U^*+\frac{4}{5}a_8\\
ReSD^{2U*} & = & \frac{2\sqrt{5}}{7}ReD^0D^{2U*}-\frac{\sqrt{15}}{14} (|D^U|^2-|D^N|^2)\\
\nonumber
ReLD^{2U*} & = & a_9-\bigl(ReD^UU^*-ReD^NN^*\bigr)\\
|D^U|^2+|D^N|^2 & = & 2|D^0|^2-\frac{4}{7}a_{10}\\
\nonumber
|D^U|^2-|D^N|^2 & = & a_{12}-\sqrt{3}ReD^0D^{2U^*}=\frac{4}{7}a_{12}-2\sqrt{\frac{3}{5}}ReSD^{2U*}\\
|D^{2U}|^2+|D^{2N}|^2 & = & 2|D^0|^2-\frac{2}{7}a_{10}\\
\nonumber
|D^{2U}|^2-|D^{2N}|^2 & = & a_{15}
\end{eqnarray}

The $D$-wave contribution $\Delta_{1,u}$ to $a_{1,u}$ is obviously not zero. Our analysis assumes that 
\begin{equation}
a_{1,u}+a_{2,u}=c_{1,u}+c_{2,u}=|S|^2+3|L|^2
\end{equation}
This approximation is necessary in order to calculate the scaling factor $K^2$ in (5.12) to determine the $S$- and $P$-wave $\tau=u$ amplitudes and thus to enable to determine the $D$-wave amplitudes themselves. To evaluate this approximation quantitatively we calculate a posteriori $\Delta_{1,u}$ for each Monte Carlo solution and determine
\begin{equation}
DRAT=\frac{\Delta_{1,u}}{a_{1,u}+a_{2,u}}
\end{equation}
where $DRAT$ is a measure of the approximation of $K^2$ compared to $(K^*)^2$ which includes the calculated $D$-waves
\begin{equation}
(K^*)^2 = \frac{(a_{1,u}+\Delta_{1,u})+a_{2,u}}{a_{1,d}+a_{2,d}}=K^2(1+DRAT)
\end{equation}
We introduced a plausible cut-off $DRAT > 0.15$. The average values of $DRAT$ depend on the mass bin and range from low 0.020 to high 0.080 with most bins at $\sim 0.060$. These values of $DRAT$ demonstrate that the approximation (6.6) is a good approximation. 

For the sake of completness we present the remaining three groups of equations
\begin{eqnarray}
ReD^UU^*+ReD^NN^* & = & \sqrt{3}\bigl(ReD^0L^*-a_7\bigr)\\
\nonumber
ReD^UU^*-ReD^NN^* & = & -ReLD^{2U*}+a_9\\
ReUD^{2U*}+ReND^{2N*} & = & 4ReD^UL^*+2\sqrt{3}ReD^0U^*-4a_8\\
\nonumber
ReUD^{2U*}-ReND^{2N*} & = & a_{13}\\
ReD^UD^{2U*}+ReD^ND^{2N*} & = & 2\sqrt{7}\bigl(ReD^UD^{0*}-a_{11}\bigr)\\
\nonumber
ReD^UD^{2U*}-ReD^ND^{2N*} & = & a_{14}
\end{eqnarray}

\subsection{Decoherence free Kraus amplitude $D^0_u$}

Evidence from the amplitude analysis Ref.~\cite{becker79b} presented in Ref.~\cite{svec13b} shows a single solution for the amplitude $D^0_\tau$ which satisfies phase conditions $\Phi(L_\tau S^*_\tau)=\Phi(L_\tau D^{0*}_\tau)-\Phi(S_\tau D^{0*}_\tau)$ for $\tau=u,d$. The amplitude $D^0$ is thus a decoherence free Kraus amplitude given by
\begin{equation}
D^0(\ell)\equiv D^0 = e^{i\psi} D^{0,0}
\end{equation}
CERN measurements below 1080 MeV at low $t$ indicate that $a_{7,\tau}=0$. The equations (6.1) then read
\begin{eqnarray}
ReD^0S^* & = & 0\\
\nonumber
ReD^0L^* & = & \frac{1}{\sqrt{5}}\Delta_4
\end{eqnarray}
Recall the spin mixing mechanism for amplitudes $S$ and $L$ 
\begin{eqnarray}
S & = & e^{i\phi} \bigl ( -\sin \theta S^0 + e^{i\phi} \cos\theta L^0 \bigr )\\
\nonumber
L & = & e^{i\phi} \bigl ( +\cos \theta S^0 + e^{i\phi} \sin\theta L^0 \bigr )
\end{eqnarray}
The relative phases $\phi(D^{0,0})-\phi(S^0)=0$ and $\phi(D^{0,0})-\phi(L^0)=0$~\cite{svec13a}. With $\alpha=\psi-\phi$ we obtain from 
$ReD^0S^*=0$
\begin{equation}
\tan \alpha = \frac{\sin \theta |S^0|-\cos \phi \cos \theta |L^0|}{\sin \phi \cos \theta |L^0|}
\end{equation}
and from the second equation in (6.13)
\begin{equation}
|D^0|=|D^{0,0}|=\frac{\frac{1}{\sqrt{5}} \Delta_4}{\cos \alpha \bigl(\cos \theta |S^0|+\cos \phi \sin \theta |L^0| \bigr) + \sin\alpha \sin \phi \sin \theta |L^0|}
\end{equation}

\subsection{Decoherence free amplitudes $D^U_u$ and $D^N_u$}

The pairs of amplitudes $U_\tau,D^U_\tau$ and $N_\tau,D^N_\tau$ form dephasing doublets (1.7) of decoherence free amplitudes. Omitting the subscript $\tau$ the spin mixing mechanism for these Kraus amplitudes for $\tau=u$ reads~\cite{svec13b}
\begin{eqnarray}
D^U(\ell) & = & V(D^UU)U^0+V(D^UD^U)D^{U,0}=e^{i\eta}D^{U,eff}\\
\nonumber
U(\ell)   & = & V(UU)U^0+V(UD^U)D^{U,0}=e^{i\delta}U^{eff}
\end{eqnarray}
where $V(AB)=V_\ell(AB)$ is a unitary dephasing matrix and where the effective amplitudes $U^{eff}$ and $D^{U,eff}$ have the same relative phase as the corresponding $S$-wave amplitudes given in the Table II of Ref.~\cite{svec13a}. For weak $\rho^0(770)$ mixing we expect $U^{eff}\sim U^0$ and $\delta=2\phi' \sim 2\phi$ (eq.(5.14)). The $\rho^0(770)$ spin mixing reveals itself as a $\rho^0(770)$ structure in $|D^{U}|^2=|D^{U,eff}|^2$. We shall seek a decoherence free solution for $D^U$ and $D^N$ using (6.2) for $ReD^US^*$ and $Re D^U L^*$. 

To render the equations (6.2) solvable we assume 
$2\sqrt{21}a_{11} - (2\sqrt{21}+1)ReD^UD^{0*}=0$ and $a_8=0$. With $D^0$ and $U$ known, we can calculate the term $Re D^0U^*$. Then the equations (6.2) read
\begin{eqnarray}
ReD^US^* & = & \Delta_{5 M} = |D^U||S|\cos \Phi(D^US^*)\\
\nonumber
ReD^UL^* & = & \Delta_{6 M} = |D^U||L|\cos \Phi(D^UL^*)
\end{eqnarray}
where $\Delta_{k M},k=5,6$ are the known terms on the r.h.s. of equations (6.2). We assume that $D^U$ is a decoherence free Kraus amplitude $D^U=D^U(\ell)$. Then the correlation factors $\cos \Phi(D^US^*)$ and $\cos \Phi(D^UL^*)$ are equal to the cosines of relative phases of the Kraus decoherence free ampltudes
\begin{eqnarray}
\cos \Phi(D^US^*) & = & \cos \Phi(D^U(\ell)S^*)\\
\nonumber
\cos \Phi(D^UL^*) & = & \cos \Phi(D^U(\ell)L^*)
\end{eqnarray}
for all $\ell=1,M$. It follows that
\begin{equation}
\cos \Phi(D^U(\ell)S^*)=\frac{|L|\Delta_{5 M} }{|S|\Delta_{6 M}}\cos \Phi(D^U(\ell)L^*)
\end{equation} 
Kraus amplitudes must satisfy cosine condition
\begin{equation}
\cos^2 \Phi(LS^*)+\cos^2 \Phi(D^US^*)+\cos^2 \Phi(D^UL^*)-
2\cos \Phi(LS^*)\cos \Phi(D^US^*)\cos \Phi(D^UL^*)=1
\end{equation}
where we have omitted the label $\ell$. With (6.20) and $c_4=|L||S|
\cos \Phi(LS^*)$ the cosine condition can be solved for 
\begin{equation}
\cos \Phi(D^U(\ell)L^*)=\frac{\Delta_{6M}|S|\sin \Phi(LS^*)}
{\sqrt{|L|^2\Delta^2_{5M}+|S|^2\Delta^2_{6M}-2c_4\Delta_{5M}\Delta_{6M}}}
\end{equation}
From (6.20) we obtain
\begin{equation}
\cos \Phi(D^U(\ell)S^*)=\frac{\Delta_{5M}|L|\sin \Phi(LS^*)}
{\sqrt{|L|^2\Delta^2_{5M}+|S|^2\Delta^2_{6M}-2c_4\Delta_{5M}\Delta_{6M}}}
\end{equation}
The physical constraints $|L|^2\Delta^2_{5M}+
|S|^2\Delta^2_{6M}-2c_4\Delta_{5M}\Delta_{6M}>0$ and $|\cos \Phi(D^US^*)|\leq 1$, $|\cos \Phi(D^US^*)|\leq 1$ impose correlated restrictions on $\Delta_{5M}$ and $\Delta_{6M}$ which define four distinct domains in the data error volume at each $(m,t)$ bin based on their values
\begin{eqnarray}
  & D(++): & \Delta_{5M}(++) \geq 0, \quad \Delta_{6M}(++)\geq 0\\
\nonumber
  & D(-+): & \Delta_{5M}(-+) < 0, \quad \Delta_{6M}(-+)\geq 0\\
  & D(+-): & \Delta_{5M}(+-) \geq 0, \quad \Delta_{6M}(+-)< 0\\
\nonumber  
  & D(--): & \Delta_{5M}(--) < 0, \quad \Delta_{6M}(--) < 0
\end{eqnarray}
It thus appears that there are 4 solutions for the cosines (6.22) and (6.23). In each domain $D(ij)$ the Kraus amplitude $D^U$ is decoherence free with two bilinear terms 
\begin{eqnarray}
ReD^U(ij)S^* & = & \Delta_{5M}(ij) = |D^U||S|\cos \Phi(D^U(ij)S^*)\\
\nonumber
ReD^U(ij)L^* & = & \Delta_{6M}(ij) = |D^U||L|\cos \Phi(D^U(ij)L^*)
\end{eqnarray}
Using (6.22) and (6.23) we find from (6.26)
\begin{equation}
|D^U(ij)|^2= \frac
{|L|^2\Delta^2_{5M}(ij)+|S|^2\Delta^2_{6M}(ij)-2c_4\Delta_{5M}(ij)\Delta_{6M}(ij)}{|S|^2|L|^2\sin^2 \Phi(LS^*)}
\end{equation}
Next we apply SMM (6.14) to $ReD^U(ij)S^*=\Delta_{5M}(ij)$ and $ReD^U(ij)L^*=\Delta_{6M}(ij)$ in (6.26) to determine the phases $\eta(ij)$ in each domain $D(ij)$. With the relative phases of the effective amplitudes equal to the relative phases of the $S$-matrix amplitudes $\phi(D^{U,0})-\phi(S^0)=-\pi$ and $\phi(D^{U,0})-\phi(L^0)=-\pi$~\cite{svec13a} and with definitions
\begin{eqnarray}
X(ij) & = & |D^U|\cos \bigl(\eta(ij)-\phi \bigr)=
|D^U|\cos \beta(ij)\\
\nonumber
Y(ij) & = & |D^U|\cos \bigl(\eta(ij)-2\phi \bigr)=
|D^U|\cos (\beta(ij)-\phi)
\end{eqnarray}
the SMM for the domain $D(ij)$ reads
\begin{eqnarray}
\Delta_{5M}(ij))= +X(ij)\sin \theta |S^0|-Y(ij)\cos \theta |L^0|\\
\nonumber
\Delta_{6M}(ij))= -X(ij)\cos \theta |S^0|-Y(ij)\sin \theta |L^0|
\end{eqnarray}
Solving for $X(ij)$ and $Y(ij)$ and using $Z(ij)=\bigl(Y(ij)-X(ij)\cos \phi\bigr)/\sin \phi$ we find 
\begin{eqnarray}
|D^{U,eff}(ij)|^2 = |D^U(ij)|^2 & = & X^2(ij)+Z^2(ij)\\
\nonumber
\tan \beta(ij) & = & \frac{Z(ij)}{X(ij)}
\end{eqnarray}
Then $\eta(ij)=\beta(ij)+\phi$. Substituting the solutions for $X(ij)$ and $Z(ij)$ into the expression for $|D^U(ij)|^2$ in (6.30) we recover (6.27) indicating the self-consistency of the calulations of the phases.

$S$- and $P$-wave amplitude analyses of the CERN data below 1080 MeV at low $t$ found 
$|U_\tau|^2 \approx |N_\tau|^2$. The measurements on unpolarized and polarized targets suggest $a_{15}=0$ which implies $|D^{2U}|^2 =|D^{2N}|^2$. These facts suggest that we may assume $|D^U|^2 =|D^N|^2$ which allows us to determine the amplitude $D^N=e^{i\eta} D^{N,eff}$.

\subsection{Decohering amplitudes $D^{2U}_u$ and $D^{2N}_u$}

First evidence for decohering amplitude $D^{2U}$ came from the analysis of CERN meausurement of $\pi^- p \to \pi^- \pi^+ n$ on polarized target at large momentum transfer $t$~\cite{rybicki85}. Figure 10 shows the violation of the phase and cosine conditions for the triplet $S_\tau,D^0_\tau,D^{2U}_\tau$. Since $S$ and $D^0$ are decoherence free amplitudes, the violations indicate that $D^{2U}$ and $D^{2N}$ are decohering amplitudes
\begin{eqnarray}
D^{2U}_\tau(\ell) & = & \exp(i\chi(\ell))D^{2U,0}_\tau\\
\nonumber
D^{2N}_\tau(\ell) & = & \exp(i\chi(\ell))D^{2N,0}_\tau
\end{eqnarray}
where
$\ell=1,M$ and $2 \leq M \leq 4$~\cite{svec13b}. Motivated by experiment we assume in the equations (6.3) that $|D^U|^2-|D^N|^2=0$ and $a_9=0$ and obtain
\begin{eqnarray}
ReSD^{2U*} & = & \frac{2\sqrt{5}}{7}ReD^0D^{2U*}\\
\nonumber
ReLD^{2U*} & = &-\bigl(ReD^UU^*-ReD^NN^*\bigr)
\end{eqnarray}
The first relation must hold true for any set of probabilities $p_{\ell \ell}$ so it must hold true also for the cosines of relative phases of the Kraus amplitudes. Then
\begin{equation}
\cos \Phi(SD^{2U*}(\ell))=\frac{2\sqrt{5}}{7}\frac{|D^0|}{|S|} \cos \Phi(D^0D^{2U*}(\ell))
\end{equation}
With $ReD^{0}S^*=0$ and (6.33) we find from the cosine condition for Kraus amplitudes $S,D^0,D^{2U}$
\begin{eqnarray}
\cos \Phi(D^0D^{2U*})^{\pm} & = & \pm \frac{7|S|^2}
{\sqrt{49|S|^2 +20|D^0|^2}}\\
\nonumber
\cos \Phi(SD^{2U*})^{\pm} & = & \pm \frac{2\sqrt{5}|D^0|^2}{\sqrt{49|S|^2 +20|D^0|^2}}
\end{eqnarray}
Using the result (6.34) for $\cos \Phi(SD^{2U*})^{\pm}$ in the cosine condition for the Kraus amplitudes $S,L,D^{2U}$ we obtain two quadratic equations for 
$X^{\pm}= \cos \Phi(LD^{2U*})^{\pm}\sqrt{49|S|^2 +20|D^0|^2}$
\begin{eqnarray}
(X^+)^2 -4\sqrt{5}|D^0|\cos \Phi(LS^*)X^+ +20|D^0|^2 -(49|S|^2 +20|D^0|^2)\sin^2 \Phi(LS^*)=0\\
\nonumber
(X^-)^2 +4\sqrt{5}|D^0|\cos \Phi(LS^*)X^- +20|D^0|^2 -(49|S|^2 +20|D^0|^2)\sin^2 \Phi(LS^*)=0
\end{eqnarray}
The solutions of (6.35) read
\begin{eqnarray}
\cos \Phi(LD^{2U*})^+_{1,2} & = & \cos \Phi(SD^{2U*})^+\cos \Phi(LS^*) \pm 
\cos \Phi(D^0D^{2U*})^{\pm} \sin \Phi(LS^*)\\
\nonumber
\cos \Phi(LD^{2U*})^-_{1,2} & = & \cos \Phi(SD^{2U*})^-\cos \Phi(LS^*) \pm 
\cos \Phi(D^0D^{2U*})^{\pm} \sin \Phi(LS^*)
\end{eqnarray}

\begin{figure} [htp]
\includegraphics[width=12cm,height=10.5cm]{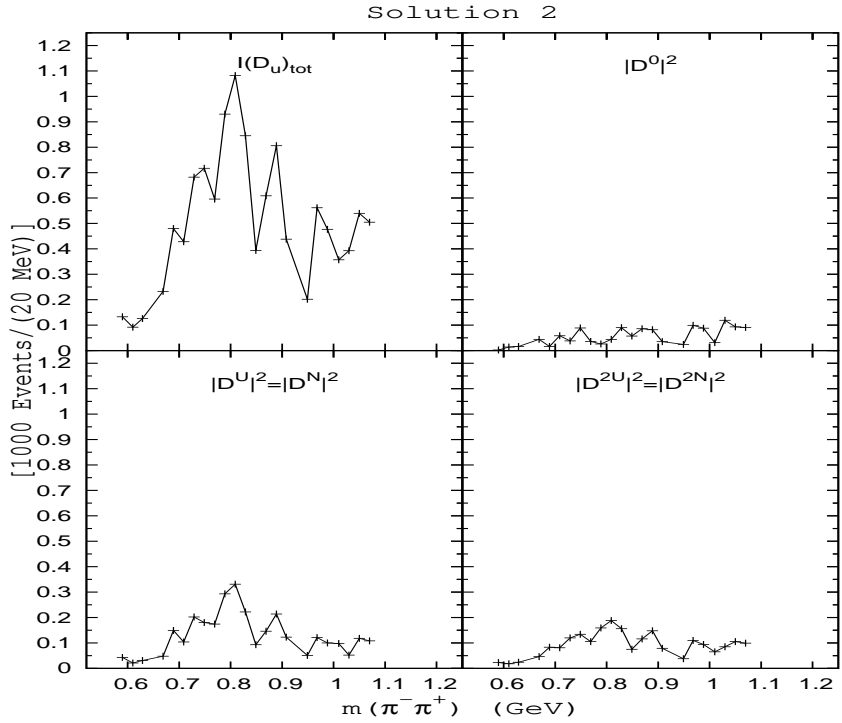}
\caption{Solution 2 for the total $D$-wave intensity $I(D_u)_{tot}$ and for the moduli $|D^0|^2$, $|D^U|^2$ and $|D^{2U}|^2$.}
\label{Figure 7}
\end{figure}

\begin{figure} [hp]
\includegraphics[width=12cm,height=10.5cm]{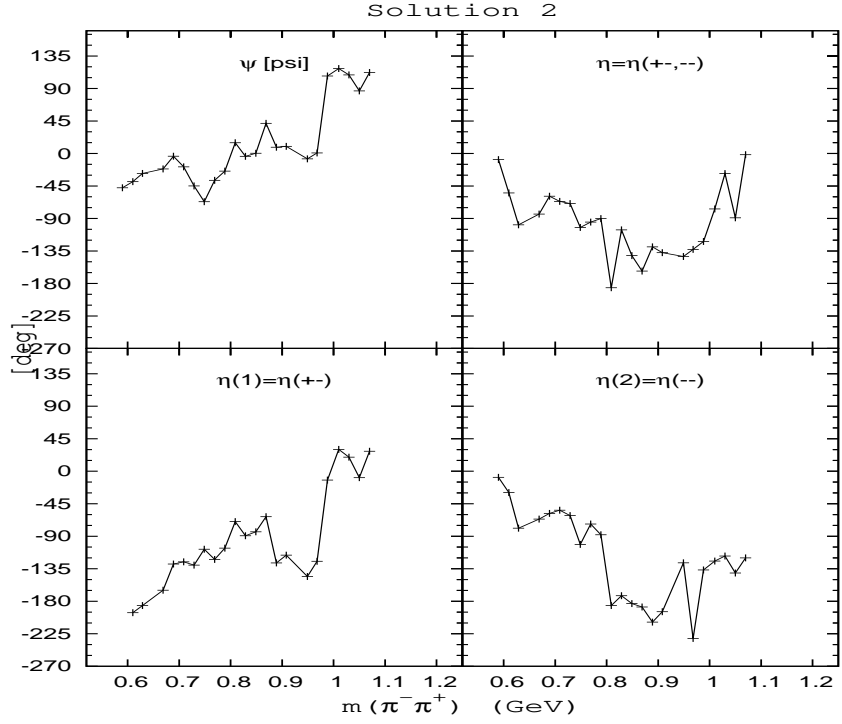}
\caption{Solution 2 for the phases $\psi$ and $\eta$ of the decoherence free amplitudes $D^0$ and $D^U$, respectively.}
\label{Figure 8}
\end{figure}

\begin{figure} [hp]
\includegraphics[width=12cm,height=10.5cm]{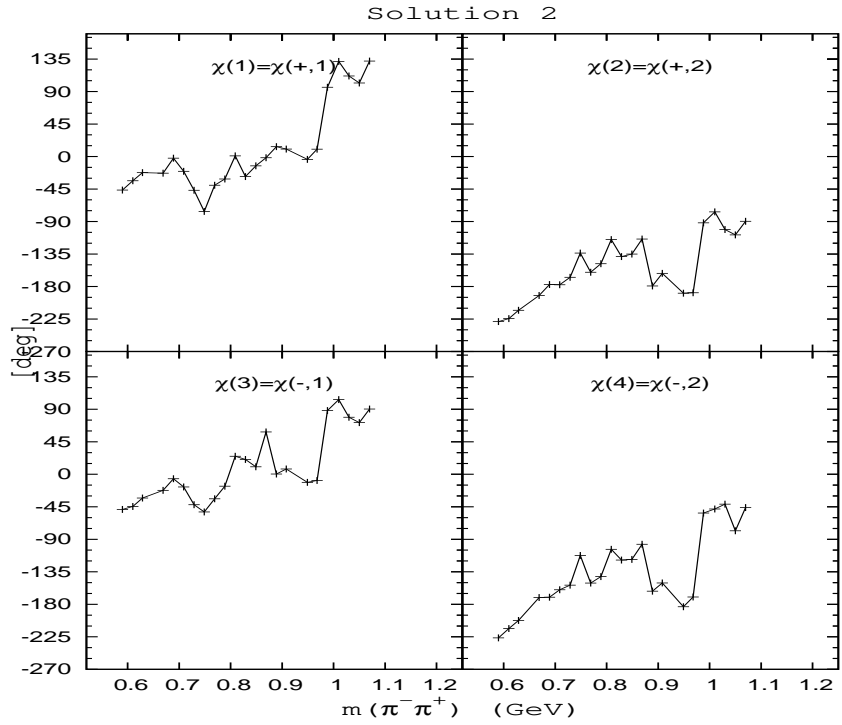}
\caption{Solution 2 for the four phases of the decohering amplitude $D^{2U}$.}
\label{Figure 9}
\end{figure}

\begin{figure} [htp]
\includegraphics[width=12cm,height=10.5cm]{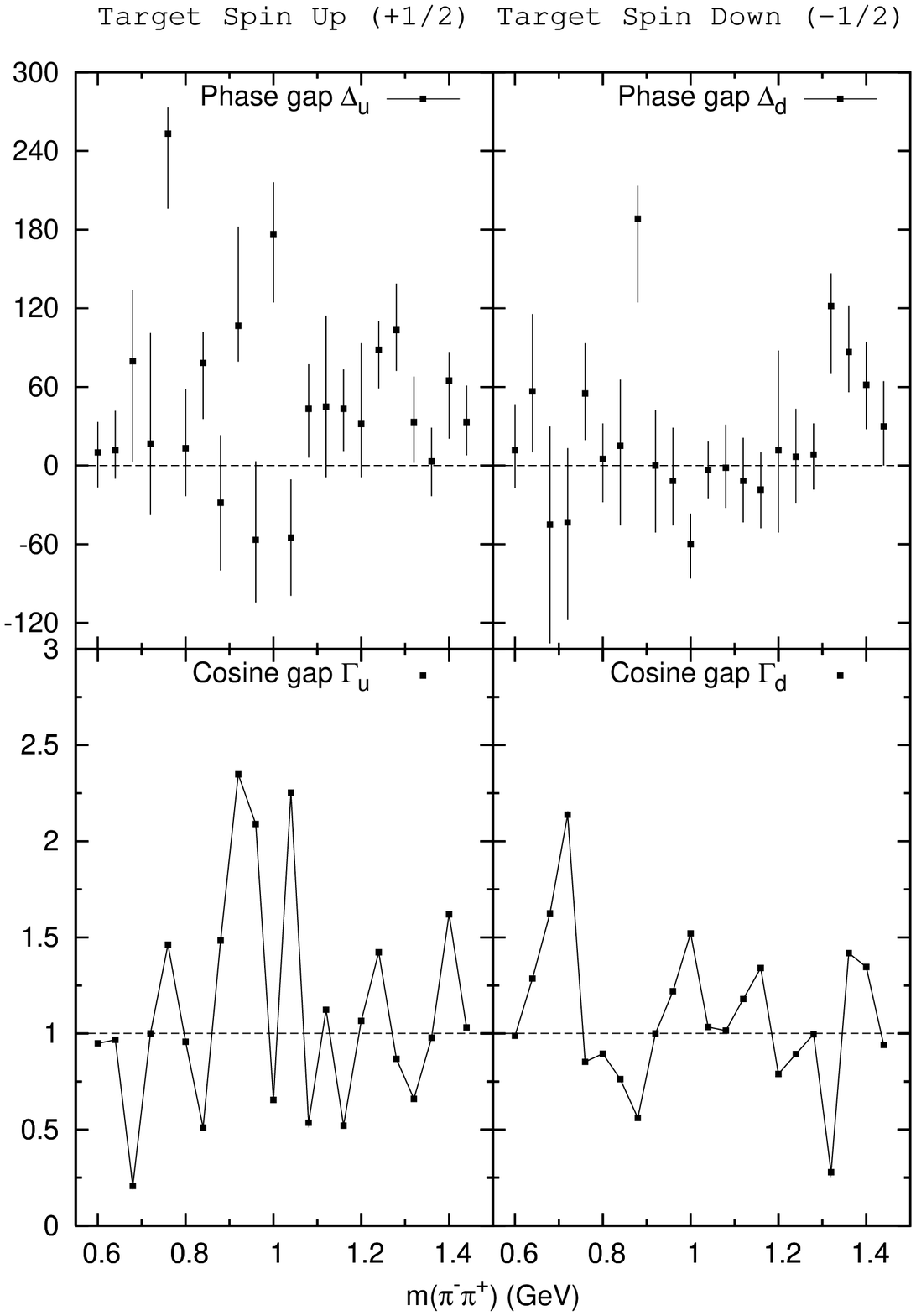}
\caption{Phase gap $\Delta_\tau$ and cosine gap $\Gamma_\tau$ for the amplitudes $L_\tau$, $D^0_\tau$ and $D^{2U}_\tau$ at large $t$. Data from~\cite{rybicki85}.}
\label{Figure 10}
\end{figure}

where
\begin{eqnarray}
\cos \Phi(LD^{2U*})^-_1 = -\cos \Phi(LD^{2U*})^+_2\\
\nonumber
\cos \Phi(LD^{2U*})^-_2 = -\cos \Phi(LD^{2U*})^+_1
\end{eqnarray}
We now define the obvious corresponding cosines
\begin{eqnarray}
\cos \Phi(SD^{2U*})^+_{1,2} & = & \cos \Phi(SD^{2U*})^+\\
\nonumber
\cos \Phi(SD^{2U*})^-_{1,2} & = & \cos \Phi(SD^{2U*})^-
\end{eqnarray}
to construct four pairs of solutions for $\cos \Phi(SD^{2U*})$ and 
$\cos \Phi(LD^{2U*})$ from which we determine four pairs of bilinear terms of the Kraus amplitudes $ReSD^{2U*}$ and $ReLD^{2U*}$. To these pairs of  bilinear terms we then apply spin mixing mechanism (6.14). First we define
\begin{eqnarray}
X(\ell) & = & \cos(\chi(\ell)-\phi)=\cos \beta(\ell)\\
\nonumber
Y(\ell) & = & \cos(\chi(\ell)-2\phi)=\cos (\beta(\ell)-\phi)
\end{eqnarray}
where $\ell=1,2,3,4$ correspond to solutions $(+,1),(+,2),(-,1),(-,2)$, respectively. Cancelling $|D^{2U}|=|D^{2U,0}|$ in the SMM relations the SMM relations take the form
\begin{eqnarray}
|S|\cos \Phi(SD^{2U*}(\ell)) & = & +X(\ell)\sin \theta |S^0|-Y(\ell)\cos \theta |L^0|\\
\nonumber
|L|\cos \Phi(LD^{2U*}(\ell)) & = & -X(\ell)\cos \theta |S^0|-Y(\ell)\sin \theta |L^0|
\end{eqnarray}
where we used $\phi(D^{2U,0})-\phi(S^0)=-\pi$ and $\phi(D^{2U,0})-\phi(L^0)=-\pi$~\cite{svec13a} for the relative phases of the $S$-matrix amplitudes. Solving for $X(\ell)$ and $Y(\ell)$ and calculating
\begin{equation}
Z(\ell)=\sin (\beta(\ell))=\frac{Y(\ell)-X(\ell)\cos \phi}{\sin \phi}
\end{equation}
we find the four phases $\chi(\ell)=\beta(\ell)+\phi$. The relations (6.39) and (6.42) imply for a single Monte Carlo solution
\begin{eqnarray}
\chi(3) & = & \chi(2)\pm\pi\\
\nonumber
\chi(4) & = & \chi(1)\pm\pi
\end{eqnarray}
These relations do not hold for the averages (mean values) of the Monte Carlo solutions at any $(m,t)$ bin since different Monte Carlo solutions have different signs of $\pi$ and there are different numbers of $+\pi$ and $-\pi$ terms.

Combining the first equation in (6.4) with the first equation in (6.5) to eliminate $a_{10}$ we find
\begin{equation}
2|D^{2U}|^2+2|D^{2N}|^2 = 2|D^0|^2+|D^{U}|^2+|D^{N}|^2
\end{equation}
With $|D^{2U}|^2=|D^{2N}|^2$ due to $a_{15}=0$ and assuming $|D^{U}|^2=|D^{N}|^2$
we calculate $|D^{2U}|^2$ from the known $|D^0|^2$ and $|D^U|^2$ using
\begin{equation}
2|D^{2U}|^2=|D^0|^2+|D^U|^2
\end{equation}
Unlike the phases $\eta(+-)$ and $\eta(--)$ the phases $\chi(\ell)$ are similar on the domains $D(+-)$, $D(--)$ and the combined domain $D(+-)+D(--)$ although they are different for each $\ell=1,4$. This similarity justifies to identify the averaged results with the unique solution for the modulus and the 4 phases of the Kraus amplitudes $D^{2U}(\ell)$.

\subsection{Results for the $D$-wave Kraus amplitudes with $\tau=u$}

We used 10 million Monte Carlo samplings of the data error volume in each $(m,t)$ bin and restricted the amplitude analysis to each domain $D(ij)$ defined in (6.24) and (6.25). These four separate analyses revealed that there are only two phases $\eta(+-)$ and $\eta(--)$ which are quite distinct. In $D(++)$ and $D(-+)$ there are no solutions in 11 out of 25 mass bins and a negligible number of Monte Carlo solutions in the remaining mass bins, indicating there are no phases $\eta(++)$ and $\eta(-+)$. There are some small variations in the $S$-and $P$-wave amplitudes and in the $D$-wave moduli in the two remaining domains that should have a unique solution. An analysis on the combined domains $D(+-)+D(--)$ calculates the suitable unique solution as the average values of the amplitudes, including the average phase $\eta$.

To summarize, in each $(m,t)$ bin we have two distinct solutions for the mean values (averages of Monte Carlo solutions) on domains $D(+-)$ and $D(--)$
\begin{eqnarray}
\Delta_{5M}(ij)_{av} & = & |D^U(ij)|_{av}|S(ij)|_{av}
\cos \bigl(\Phi(D^U(ij)S^*)_{av}\bigr)\\
\nonumber
\Delta_{6M}(ij)_{av} & = & |D^U(ij)|_{av}|L(ij)|_{av}
\cos \bigl(\Phi(D^U(ij)L^*)_{av}\bigr)
\end{eqnarray}
and a single solution on the combined domain $D(+-)+D(--)$
\begin{eqnarray}
\Delta_{5M,av} & = & |D^U|_{av}|S|_{av}\cos \bigl(\Phi(D^US^*)_{av}\bigr)\\
\nonumber
\Delta_{6M,av} & = & |D^U|_{av}|L|_{av}\cos \bigl(\Phi(D^UL^*)_{av}\bigr)
\end{eqnarray}
Since there is no reason to select one of the two solutions $\eta(+-)$ and $\eta(--)$ in (6.45) we select the average solution $\eta$ in (6.46) as the physical solution. As a result our entire analysis is done on the combined domain $D(+-)+D(--)$ and it is those results that are reported in this work. In Figures 7,8,9 we present only the results for the mean values of the Monte Carlo solutions for the $D$-wave amplitudes. 

In Figure 7 we present the total contribution of the $\tau=u$ $D$-waves $I(D_u)_{tot}=|D^0|^2+|D^U|^2+|D^N|^2+|D^{2U}|^2+|D^{2N}|^2$ and the moduli of the $D$-wave amplitudes. The total $D$-wave intensity shows a clear $\rho^0(770)$ peak originating in the spin mixing amplitudes $D^U$ and $D^N$ the moduli of which show evidence of $\rho^0(770)$ mixing. In contrast, there is no evidence of such mixing in the amplitudes $D^0$, $D^{2U}$ and $D^{2N}$. Our evidence for spin mixing of $U_\tau,D^U_\tau$ and $N_\tau,D^N_\tau$ at small $t$ is in tension with no such evidence from a differnt kind of analysis at large $t$~\cite{rybicki85}. As well, we find no evidence of a $2^{++}(840)$ resonance found in this large $t$ analysis in the amplitude $D^{2U}$ but not in any other $D$-wave amplitude~\cite{rybicki85}.

Figure 8 shows the phase $\psi$ of the amplitude $D^0$ and the  averaged/decoherence free phase $\eta$ of the amplitude $D^U$. For a comparison we show in Figure 8 also the two phases $\eta(+-)$ and $\eta(--)$. Figure 9 shows the four phases of the decohering amplitude amplitude $D^{2U}$ for $\ell=1,M$. On the basis of this analysis we conclude that the number of interacting degrees of freedom of the quantum environment is $M=4$.

Consider an observed bilinear term of $D^{2U}=D^{2U}_\tau$ with a decoherence free amplitude $A=L_\tau$ or $A=D^0_\tau$
\begin{equation}
|D^{2U}||A|\cos \Phi(D^{2U}A^*)=|D^{2U}||A|\sum \limits_{\ell=1}^4 p_{\ell \ell} \cos (\chi(\ell)-\phi(A))
\end{equation}
Each event $\pi^- p \to \pi^- \pi^+ n$ interacts with a single quantum state $\rho(E)$ with specific values of diagonal terms $p_{\ell \ell}$. The probabilities $p_{\ell \ell}$ are different for each event and the measured bilinear terms correspond to their averaged values which are somewhat different in each $(m,t)$ bin. The two pairs of phases $\chi(1),\chi(3)$ and $\chi(2),\chi(4)$ differ approximately by 180$^\circ$. As a function of the dipion mass each phase changes by about 180$^\circ$ from the low to high values of mass. The phase $\phi(D^0)=\psi$ shows a similar behavior while the phase $\phi(L)=\Phi(LS^*)$ varies slowly. From this behavior of the phases with fluctuating $p_{\ell \ell}$ we expect large fluctuations of the correlation factor from bin to bin. Thus our results for the phases $\chi(\ell)$ predict not only a violation of the phase and cosine conditions, but large fluctuations in the phase and cosine gaps $\Delta_\tau$ and $\Gamma_\tau$, which are the deviations from the theoretical  values $\Delta_\tau(th)=0$ and $\Gamma_\tau(th)=1.0$. Figure 10 shows the experimental results for the triplet $L_\tau,D^0_\tau,D^{2U}_\tau$ from the analysis at large $t$~\cite{rybicki85}. The observed large fluctuations confirm our expectations. Recall that $\chi(\ell)$ do not depend on $t$. The large $t$ analysis on its own means that $M>1$. Thus both analyses agree that the amplitudes $D^{2U}$ and $D^{2N}$ are decohering amplitudes.

\section{Spin mixing and $S$-Matrix helicity amplitudes.}

Helicity amplitudes $A^J_{\lambda \chi,0 \nu}$ with definite $t$-channel naturality were defined and related to transversity amplitudes of definite $t$-channel naturality in Ref.~\cite{lutz78,svec13a} for any dipion spin $J$ and helicity $\lambda$. Due to the $P$-parity conservation only helicity nonflip and helicity flip amplitudes $A^J_{\lambda,0}=A^J_{\lambda +,0 +}$ and $A^J_{\lambda,1}=A^J_{\lambda +,0 -}$ are independent. Here $n=0,1$ is nucleon helicity flip $n=|\chi-\nu|$. The helicity amplitudes $A_n$ are related to the transversity amplitudes $A_\tau$ by 
relations~\cite{lutz78,svec13a}
\begin{equation}
A_n=\frac{(-i)^n}{\sqrt{2}}(A_u+(-1)^n A_d)
\end{equation}
It is convenient to introduce reduced transversity amplitudes
\begin{eqnarray}
A_u & = & A \exp {i \Phi(S_u)}\\
\nonumber
A_d & = & \overline {A} \exp {i \omega} \exp {i \Phi(S_u)}
\end{eqnarray}
where  $\Phi(S_u)$ is the arbitrary absolute phase and
\begin{equation}
\omega=\Phi(S_d)-\Phi(S_u)
\end{equation}
is the relative phase between $S$-wave amplitudes of opposite transversity. The phases of $A$ and $\overline{A}$ are $\Phi_{AS}=\Phi(A_u)-\Phi(S_u)$ and $\overline{\Phi}_{AS}=\Phi(A_d)-\Phi(S_d)$, respectively. In terms of the reduced transversity amplitudes we can write the helicity amplitudes in the form
\begin{equation}
A_n=\frac{(-i)^n}{\sqrt{2}} \bigl(A+(-1)^n \overline{A}
\exp(i \omega)\bigr)\exp(i\Phi(S_u))
\end{equation}
For the moduli we find
\begin{equation}
|A_n|^2 = \frac{1}{2} \Bigl (|A|^2+|\overline{A}|^2+(-1)^n2X_A \cos(\omega)+
(-1)^n2Y_A \sin(\omega) \Bigr) 
\end{equation}
where $X_A=Re(A \overline{A}^*)$, $Y_A=Im(A \overline{A}^*)$. Note that $Y_S=Im(S \overline{S}^*)=0$ as both $S$ and $\overline{S}$ are real and positive. For the bilinear terms $A_nB_n^*$ we obtain
\begin{equation}
A_nB_n^*={1\over{2}} \Bigl ( AB^* + \overline{A}\overline{B}^* +(-1)^n \bigl (  A \overline{B}^* e^{-i\omega} +\overline{A}B^* e^{+i\omega} \bigr) \Bigr)
\end{equation}
The phase $\omega$ can be determined analytically from the consistency condition~\cite{svec12a}
\begin{equation}
|A_n|^2|B_n|^2=(Re(A_nB_n^*))^2+(Im(A_nB_n^*))^2
\end{equation}
Previous amplitude analysis of the $S$- and $P$-wave subsystem determined that 
$\cos \omega=-1$~\cite{svec12a}. Recall that $|A|=|A_u|$ and $|\overline{A}|=|A_d|$. Then the moduli of the $S$- and $P$-wave helicity amplitudes in terms of the known transversity amplitudes read
\begin{eqnarray}
|S_n|^2 & = & \frac{1}{2}\bigl(|S_u|^2+|S_d|^2-(-1)^n2|S_u||S_d| \bigr)\\
\nonumber
|L_n|^2 & = & \frac{1}{2}\bigl(|L_u|^2+|L_d|^2-(-1)^n2|L_u||L_d|
\cos (\Phi_{LS}-\overline{\Phi}_{LS}) \bigr)\\
\nonumber
|U_n|^2 & = & \frac{1}{2}\bigl(|U_u|^2+|U_d|^2-(-1)^n2|U_u||U_d|
\cos (\Phi_{US}-\overline{\Phi}_{US}) \bigr)\\
\nonumber
|N_n|^2 & = & \frac{1}{2}\bigl(|N_u|^2+|N_d|^2-(-1)^n2|N_u||N_d|
\cos (\Phi_{NS}-\overline{\Phi}_{NS})\bigr)
\end{eqnarray}
\begin{figure} [htp]
\includegraphics[width=12cm,height=10.5cm]{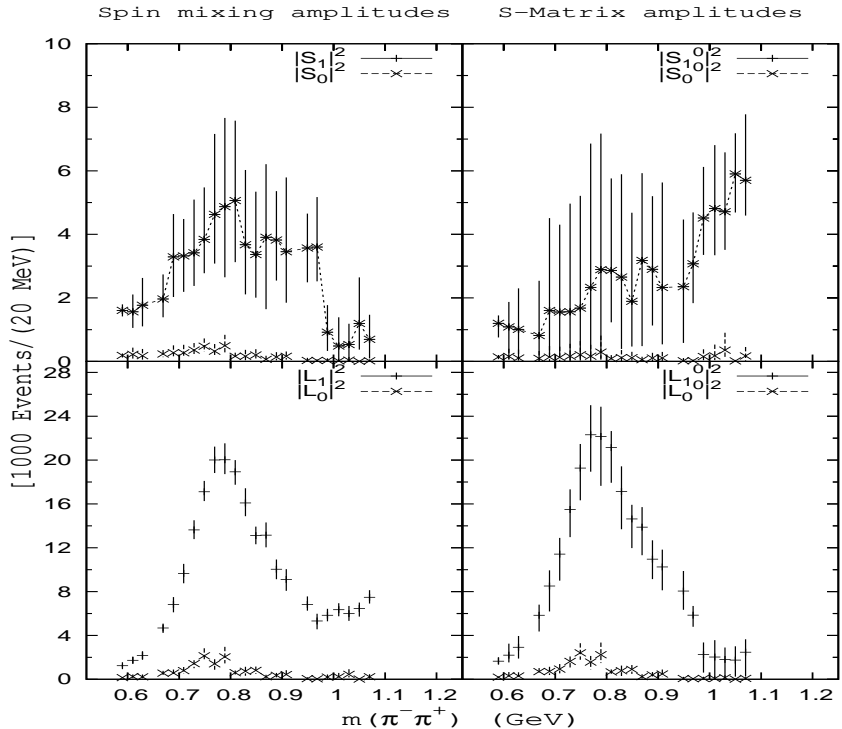}
\caption{Spin mixing and $S$-matrix helicity amplitudes $|S_n|^2,|L_n|^2$ and $|S^0_n|^2,|L^0_n|^2$.}
\label{Figure 11}
\end{figure}
From the nalysis with SMM we get $\Phi_{LS}=\overline{\Phi}_{LS}$. At small $t$ the non-flip amplitude $N_0$ dominates the single flip amplitude $N_1$ due to the $a_2$ exchange which allows us to set $\cos (\Phi_{NS}-\overline{\Phi}_{NS}) = -1$ since $|N_u|^2 \approx |N_d|^2$. Similar relations hold true for the moduli of the $S$-matrix helicity amplitudes. We assume that $\omega^0=\Phi(S^0_d)-\Phi(S^0_u)=\omega$. Taking into account the self-consistent relative phases of the $S$-matrix transversity 
amplitudes~\cite{svec13a} $\Phi_{L^0S^0}=\overline{\Phi}_{L^0S^0}=0$ and $\Phi_{U^0S^0}=\overline{\Phi}_{U^0S^0}=-\pi$ and the relations $|U_\tau|=|U^0_\tau|$, 
$|N_\tau|=|N^0_\tau|$, we find
\begin{eqnarray}
|S^0_n|^2 & = & \frac{1}{2}\bigl(|S^0_u|-(-1)^n|S^0_d|\bigr)^2\\
\nonumber
|L^0_n|^2 & = & \frac{1}{2}\bigl(|L^0_u|-(-1)^n|L^0_d| \bigr)^2\\
\nonumber
|U^0_n|^2 & = & |U_n|^2\\
\nonumber
|N^0_n|^2 & = & |N_n|^2
\end{eqnarray}

It is instructive to calculate the bilinear terms $ReL_nS_n^*$ and $ImL_nS_n^*$ from (7.6). For spin mixing amplitudes we find
\begin{eqnarray}
ReL_nS_n^* & = & \frac{1}{2}\cos \Phi_{LS}(|L_u|-(-1)^n|L_d|)(|S_u|-(-1)^n|S_d|)\\
\nonumber
ImL_nS_n^* & = & \frac{1}{2}\sin \Phi_{LS}(|L_u|-(-1)^n|L_d|)(|S_u|-(-1)^n|S_d|)
\end{eqnarray}
Similar relation holds for the $S$-matrix helicity amplitudes with $\Phi_{L^0S^0}=0$. From these relations follow relations for the relative phases
\begin{eqnarray}
\Phi(L_n)-\Phi(S_n) & = & \Phi_{LS}\\
\nonumber
\Phi(L^0_n)-\Phi(S^0_n) & = & 0
\end{eqnarray}

Our results for the spin mixing and $S$-matrix helicity amplitudes $|S_n|^2,|L_n|^2$ and $|S^0_n|^2,|L^0_n|^2$ are shown in the Figure 11. All non-flip amplitudes are very small. The $S$-wave amplitudes are very dissimilar. While the spin mixing amplitude $|S_1|^2$ shows a clear peak at $\rho^0(770)$ mass there is no such structure in the amplitude $|S^0_1|^2$. While $|S_1|^2$ dips near $f_0(980)$ mass, there is a rapid rise in $|S^0_1|^2$ above the $K\bar{K}$ threshold. The amplitudes $|L_1|^2$ and $|L^0_1|^2$ are both resonating similarly at $\rho^0(770)$. However the spin mixing dip at $f_0(980)$ mass seen in $|L_1|^2$ is absent in the amplitude $|L^0_1|^2$. The absence of any evidence for the spin mixing in the amplitudes $|S^0_1|^2$ and $|L^0_1|^2$ is consistent with their interpretation as $S$-matrix amplitudes.

\section{$\pi\pi$ phase-shift analysis below the $K\bar{K}$ threshold.}

\subsection{Determination of the phase-shifts $\delta^0_S$ and $\delta_P$}

The formalism of the $\pi\pi \to \pi\pi $ scattering and its connections to $\pi N \to \pi \pi N$ processes is well known~\cite{martin76,petersen77}. High statistics CERN-Munich data on $\pi^- p \to \pi^- \pi^+ n$ at 17.2 GeV/c on unpolarized target~\cite{grayer74} were analysed using several methods to determine $\pi\pi$ phase- 
shifts~\cite{hyams73,estabrooks73,estabrooks74,bugg96}. $\pi\pi$ phase-shift analysis using  CERN-Munich-Cracow data on $\pi^- p \to \pi^- \pi^+ n$ at 17.2 GeV/c on polarized target was reported in Ref.~\cite{kaminski97}. In these analyses model dependent methods were used to extract the single flip helicity amplitudes from the data which were then related to $\pi \pi$ scattering amplitudes using pion exchange dominance approximation. 

In our amplitude analysis the spin mixing helicity amplitudes $|S_1|^2$ and $|L_1|^2$ are model independent while the $S$-matrix helicity amplitudes $|S^0_1|^2$ and $|L^0_1|^2$ are only weakly dependent on the assumption $\omega^0=\omega$. Our aim is to determine the average phase shifts $\delta_P$ and $\delta^0_S$ below $K\bar{K}$ threshold for both sets of the helicity amplitudes. In our $\pi\pi$ phase-shift analysis we follow closely the method of Estabrooks and Martin~\cite{estabrooks74} and compare our results with their results. Since our data on $|S_1|^2$ and $|L_1|^2$ were obtained by a method significantly different from the one used by Estabrooks and Martin such comparison can be informative.

Elementary pion exchange contribution to the single-flip helicity amplitudes $S_1$ and $L_1$ is parametrized for the spin mixing and $S$-matrix amplitudes by the same form
\begin{eqnarray}
S_1 & = & Ke^{i\theta_S}C_S \frac{m}{\sqrt{q}}f_S(m)\\
\nonumber
L_1 & = &  Ke^{i\theta_P}\sqrt{3}\frac{m}{\sqrt{q}}f_P(m)
\end{eqnarray}
where for $t$-channel dipion helicity~\cite{estabrooks74}
\begin{equation}
K=N\frac{\sqrt{-t_{av}}}{\mu^2-t_{av}}|F(t_{av}|=
  N\frac{\sqrt{-t_{av}}}{\mu^2-t_{av}}e^{b(t_{av}-\mu^2)}
\end{equation}
is the overall normalization factor at a single value of the momentum transfer $t=t_{av}=0.068$ (GeV/c)$^2$ corresponding to the bin $0.005 < |t| <0.20$ (GeV/c)$^2$, $\mu$ is the pion mass and $m$ and $q=0.5\sqrt{m^2-4\mu^2}$ are the dipion mass and cms momentum, respectively. The phases $\Phi(S_1)=\theta_S+\Phi(f_S)$ and $\Phi(L_1)=\theta_P+\Phi(f_P)$ reflect the fact that experimentally $\Phi_{L_1S_1}=\Phi(L_1)-\Phi(S_1) \neq \Phi_{PS}=\Phi(f_P)-\Phi(f_S)$ due to their different origins: while $\Phi_{L_1S_1}$ arises from the amplitude analysis of the complete pion production data on polarized target the relative phase $\Phi_{PS}$ shall arise from the moduli of $|S_1|^2$ and $|L_1|^2$.
The correction factor $C_S$ is introduced in our analysis of the spin mixing amplitudes to normalize $|f_S|^2$ to the value of $|f_S(EM)|^2$ at $m=789$ MeV from the Estabrooks-Martin analysis. There is no correction in our analysis of the $S$-matrix amplitudes and $C_S=1.0$. The factor $\sqrt{3}=\sqrt{2J+1}$ for $J=1$.

In terms of $\pi\pi$ scattering amplitudes $f^I_L$ with definite isospin $I$ the amplitudes $f_S$ and $f_P$ read
\begin{eqnarray}
f_S & = & \frac{2}{3}f^0_S+\frac{1}{3}f^2_S\\
\nonumber
f_P & = & f^1_P
\end{eqnarray}
Following the Estabrooks-Martin analysis we assume elastic $\pi^- \pi^+$ scattering below $K\bar{K}$ threshold
\begin{equation}
f^I_L=\sin \delta^I_L e^{i\delta^I_L}
\end{equation}
We determine the normalization factor $K$ from the condition that $\delta^1_P=90^\circ$ at the peak value $|L_1^*|^2$ of $|L_1|^2$. Then the $P$-wave amplitude reads
\begin{equation}
|f_P|^2= \frac{q}{q^*} \frac{m^{*2}}{m^2} \frac{|L_1|^2}{|L_1^*|^2}=\sin^2 \delta^1_P 
\end{equation}
The $S$-wave amplitude is given by
\begin{equation}
|f_S|^2=\frac{q}{m^2}\frac{|S_1|^2}{K^2C_S^2}
=\frac{4}{9}|f^0_S|^2+\frac{1}{9}|f^2_S|^2+\frac{4}{9}|f^0_S||f^2_S|
        \cos (\delta^0_S-\delta^2_S)
\end{equation}
The equation (8.6) is a quadratic equation for $\sin^2 \delta^0_S$ with two solutions
\begin{equation}
(\sin^2 \delta^0_S)_{1,2}=\frac{1}{2A} \Bigl( B\pm \sqrt{B^2-AC^2} \Bigr)
\end{equation}
where
\begin{eqnarray}
A & = & 4(1+\sin^2 \delta^2_S)^2 + \sin^2 2\delta^2_S\\
\nonumber
B & = & 2C(1+\sin^2 \delta^2_S) + \sin^2 2\delta^2_S\\
\nonumber
C & = & 9|f_S|^2-\sin^2 \delta^2_S
\end{eqnarray}
For the phase-shifts $\delta^2_S$ we take the values from the Table 1 of Ref.~\cite{estabrooks74}. For $|L_1|^2,|S_1|^2$ and $|L^0_1|^2,|S^0_1|^2$ we take the average experimental values only.

Our results for $\delta^1_P$ are very close to the EM results for both spin mixing and $S$-matrix amplitudes and are not shown. Figure 12 shows our results for the phase-shifts $\delta^0_S$ and the relative phases $\Phi_{PS}$ for the spin mixing amplitudes compared to the results of Estabrooks-Martin (EM). Our key observation is that in both Solutions 1 and 2 the phase-shift $\delta^0_S$ passes through $90^\circ$ near $\rho^0(770)$ mass which reflects the $\rho^0(770)$ mixing in the $S$-wave production amplitude. Our Solution 1 (sign + in (8.7)) comes closest to the EM Solution. Above 800 MeV it rises faster than the EM solution reflecting a more spin pronounced mixing in our $|S_1|^2$ compared to a broader structure of that amplitude in EM analysis. The most pronounced signature of $\rho^0(770)$ mixing is in the Solution 2 with its steep passage through $90^\circ$ resembling the $UP$ Solution of the 
phase-shift analyses on unpolarized target. 

Figure 13 shows our results for the phase-shifts $\delta^0_S$ and the relative phases $\Phi_{PS}$ for the $S$-matrix amplitudes compared to EM results. The two Solutions for $\delta^0_S$ are very similar and neither Solution shows evidence of $\rho^0(770)$ mixing. The phase-shifts are relatively small and essentially flat below $K\bar{K}$ threshold. Their approximate equality suggests a unique physical solution for the $\pi\pi$ scattering phase shifts consistent with unitarity is attainable from the data on polarized target. In contrast, the spin mixing amplitudes yield two distinct resonating solutions for $\delta^0_S$ that must violate unitarity~\cite{pennington73} which reveals that these helicity amplitudes are not $S$-matrix amplitudes. Our results meet expectations that there can be no spin mixing in the $S$-matrix theory and confirm that the observed spin mixing must arise from a new non-unitary interaction of the produced $S$-matrix final state in $\pi^- p \to \pi^- \pi^+ n$ with a quantum environment.

\begin{figure} [htp]
\includegraphics[width=12cm,height=10.5cm]{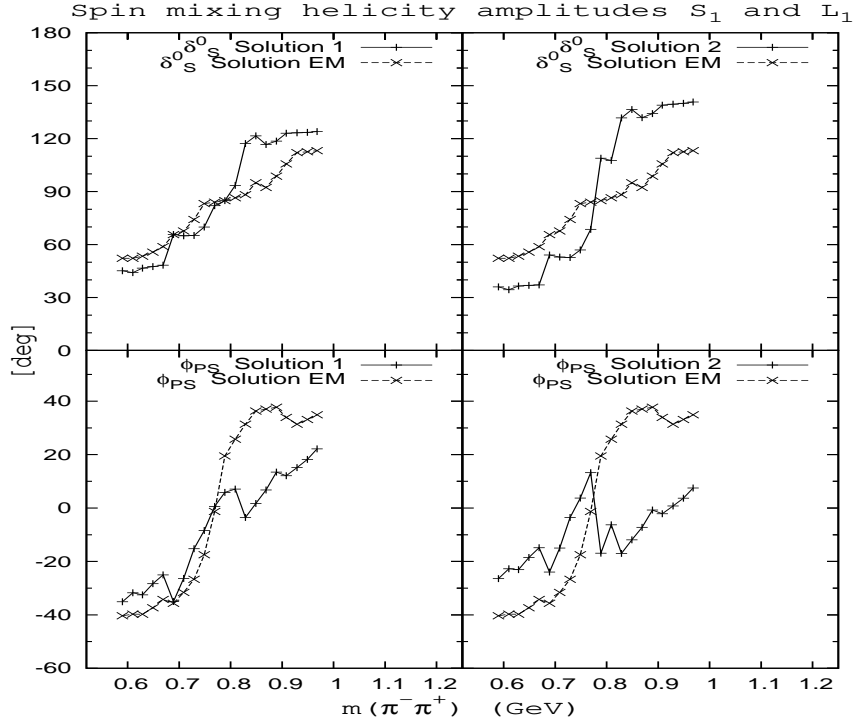}
\caption{Phase shifts $\delta^0_S$ and relative phases $\Phi_{PS}$ from the spin mixing helicity amplitudes $|S_1|^2$ and $|L_1|^2$.}
\label{Figure 12}
\end{figure}

\begin{figure} [hp]
\includegraphics[width=12cm,height=10.5cm]{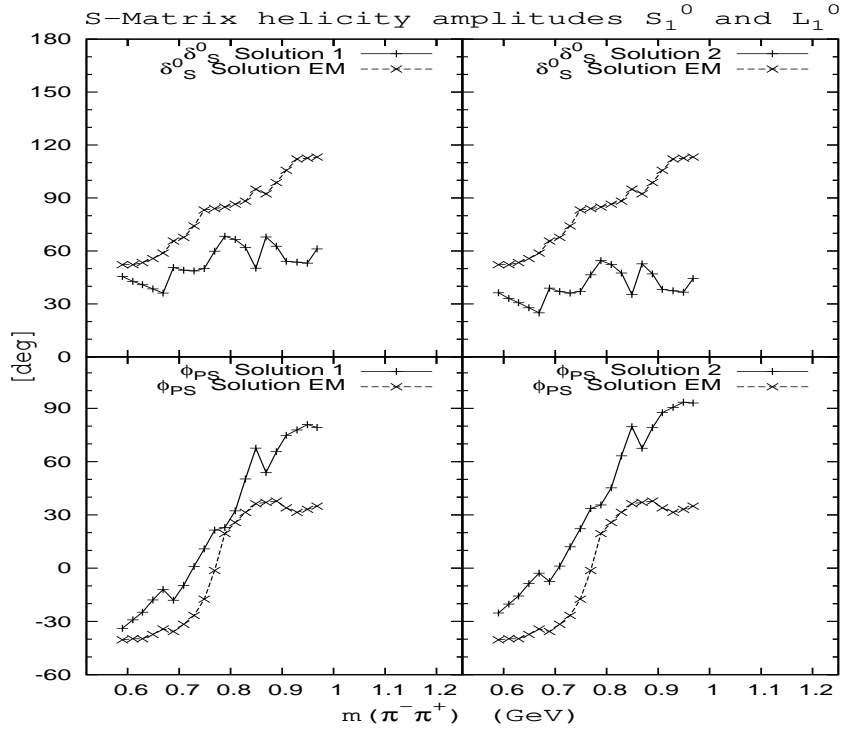}
\caption{Phase shifts $\delta^0_S$ and relative phases $\Phi_{PS}$ from the $S$-matrix  helicity amplitudes $|S^0_1|^2$ and $|L^0_1|^2$.}
\label{Figure 13}
\end{figure}

\subsection{A comparison with the Cracow phase-shift analysis}

In 1997 Kami\'{n}ski, Le\'{s}niak and Rybicki published the first phase-shift analysis using their new amplitude analysis of the CERN data on polarized target~\cite{kaminski97}. In Section VI of Ref.~\cite{svec12d} we present a survey of $S$-wave moduli and intensities of all amplitude analyses of the five measurements on polarized targets. In that survey the Figure 11 shows that their two solutions for the moduli of the transversity amplitudes $|S_u|^2$ and $|S_d|^2$ are very close to our recent results~\cite{svec12a} presented in Figure 12 with $|S_d|^2$ showing evidence of spin mixing in both analyses. However we differ in the relative phases $\Phi(L_u S^*_u)$ and $\Phi(L_d S^*_d)$. Using a Monte Carlo method we found that these relative phases shown in Figure 17 of Ref.~\cite{svec12d} do not change signs. Using a $\chi^2$ fitting method Kamiski et al. found zeros in both relative phases in both solutions near $700-800$ MeV. This enables them to claim a change of signs of these phases shown in Figure 16 of Ref.~\cite{svec12d} that closely resembles that of the relative phase $\Phi_{PS}$ from $\pi\pi$ phase shift analyses on unpolarized targets. While in our phase-shift analysis these phases do not play a direct role they are important in their phase-shift analysis. 

In terms of our notation they assume
\begin{eqnarray}
f_S & = & N_S f \frac{\sqrt{q}}{m} \bigl( a_1 S_u +a_2 S_d)\\
\nonumber
f_P & = & N_P |A_{BW}(\rho^0)|e^{i\phi(\rho^0)}
\end{eqnarray}
where $N_S$ and $N_P$ are normalization factors, $a_1,a_2$ are complex kinematical constants, $f$ is a correction factor, $A_{BW}(\rho^0)$ is a Breit-Wigner amplitude at $\rho^0(770)$ and $\phi(\rho^0)$ is the its phase.
The constants satisfy $|a_1|^2+|a_2|^2=1$ so that
\begin{eqnarray}
a_1 & = & \cos \alpha e^{i\theta_1}\\
\nonumber
a_2 & = & \sin \alpha e^{i\theta_2}
\end{eqnarray}
This compares with our $N'_S=\frac{1}{K}$, $f'=\frac{1}{C_S}$ and
\begin{eqnarray}
a'_1 & = & \frac{1}{\sqrt{2}} e^{-i\theta_S-i\frac{\pi}{2}}\\
\nonumber
a'_2 & = &  \frac{1}{\sqrt{2}} e^{-i\theta_S+i\frac{\pi}{2}}
\end{eqnarray}
They assume that the absolute phases of the transversity amplitudes are given by
\begin{eqnarray}
\Phi(S_u) & = & \Phi(S_uL_u^*)+\phi(\rho^0)\\
\nonumber
\Phi(S_d) & = & \Phi(S_dL_d^*)+\phi(\rho^0)+\Delta
\end{eqnarray}
where $\Delta$ is a correction factor. Then $|f_S|^2$ will depend on the relative phase 
\begin{equation}
\omega = \Phi(S_d)-\Phi(S_u)=\Phi(S_dL_d^*)+\Delta-\Phi(S_uL_u^*)
\end{equation}
In Ref.~\cite{svec12a} we have shown that $\omega$ can be determined analytically from a self-consistency condition of the bilinear terms of transversity amplitudes. The only solution consistent with resonant $|L_1|^2$ and pion exchange dominance of $|S_1|^2$ requires $\cos \omega=-1$, or $\omega = \pm \pi$. Apart from the zero structure of the phases $\Phi(S_dL_d^*)$ and $\Phi(S_uL_u^*)$ near 800 MeV the two amplitude analyses are very similar. The main difference then are the equations (8.10) and (8.13) which define the amplitude $f_S$ differently from our definition. Apart from $C_S$ there are no other adjustable parameters in our definition of $f_S$ while the Cracow definition (8.9) involves in addition to $f$ the adjustable parameters $a_1, a_2$ and $\Delta$. 

There are four solutions for $\delta^0_S$ in the Cracow analysis called "down-flat"(DF), "up-flat"(UF), "down-steep"(DS) and "up-steep"(US). The Solution "down-flat" $\delta^0_S(DF)$ fits best the CERN-Munich phase shift $\delta^0_S(CM)$ from the analyses on unpolarized target~\cite{grayer74}. It passes through $90^\circ$ at 770 MeV, is 
$15^\circ-10^\circ$ bellow $\delta^0_S(CM)$ for masses bellow 700 MeV and 
$10^\circ-20^\circ$ above $\delta^0_S(CM)$ for masses above 770 MeV. The Solution "up-flat" rises higher $10^\circ-100^\circ$ above $\delta^0_S(CM)$ for masses above 700 MeV. The Solutions "down-steep" and "up-steep" are similar. They both pass through $90^\circ$ at 770 MeV and rise rapidly to $\sim 140^\circ$ at 790 MeV.

The Cracow phase-shift analysis is to be compared with our phase-shift analysis of the spin mixing helicity amplitudes shown in Figure 12 since the two analyses involve the same transversity amplitudes. Our Solution 1 $\delta^0_S(1)$ is similar to the Solution "down-flat" bellow 830 MeV. For masses 830 -930 MeV it is $20^\circ-10^\circ$ higher than the Solution "down-flat" but well bellow the Solution "up-flat" so that in this mass interval
\begin{equation}
\delta^0_S(CM) < \delta^0_S(DF) < \delta^0_S(1) < \delta^0_S(UF)
\end{equation}
Above 930 MeV the Solution 1 and the Solution "down-flat" are again similar. The Solution 2 $\delta^0_S(2)$ is very similar to the nearly equal Solutions "down-steep" and "up-steep". We conclude that the two phase-shift analyses of polarized target data are mutually consistent with the principal difference occuring in our Solution 1 in the mass interval 830 -930 MeV.

\section{Quantum environment: a new view of dark matter.}

\subsection{Dark neutrino hypothesis}

The question arises - what is the nature of the quantum environment? The consistency of the pure dephasing interaction with the Standard Model~\cite{svec13b} suggests that it is a universal environment in the entire Universe which may manifest itself in some astrophysical observations. Astrophysical  observations provide a convincing evidence for the existence of dark matter and dark energy which are omnipresent environments in the Universe. Dark matter is characterized by non-standard interactions with baryonic matter. Quantum environment can be viewed as a sea of quantum states $\rho(E)$ across the Universe defined by (1.2). The four orthonormal eigenstates $|e_i>, i=1,4$ in (1.2) do not participate in any of the fundamental interactions of the Standard Model. They participate only in the non-standard pure dephasing interactions with quantum states $\rho_f(S)$ produced in particle scattering. In this aspect there is an obvious similarity between the dark matter and the quantum environment.

In priciple the eigenstates $|e_k>$ could represent some new interacting degrees characterising the dark matter and its non-standard interactions with baryonic matter. But particle physics already knows of physical eigenstates with non-standard interactions: neutrino mass eigenstates. Similarly to the eigenstates $|e_k>$, the four neutrino mass eigenstates $|m_i>,i=1,4$ do not participate in any of the fundamental interactions of the Standard Model (neutrino magnetic moment involves non-standard electromagnetic interaction~\cite{langacker10,broggini12}). Here we anticipate the fourth light sterile neutrino $\nu_s$ based on the shared dimension $M=4$. This analogy leads us to identify the eigenstates $|e_i>$ with the neutrino mass eigenstates $|m_i>$. Flavour neutrinos are pure states 
\begin{equation}
\rho(\nu_\alpha)=|\nu_\alpha><\nu_\alpha|=\sum \limits_{i,j=1}^4 p_{ij}(\alpha)|m_i><m_j|
\end{equation}
with fixed matrix elements given by the mixing matrix $p_{ij}(\alpha)=U_{\alpha i}U^*_{\alpha j}$. The quantum states $\rho(E)$ are mixed states 
\begin{equation}
\rho(E)=\sum \limits_{i,j=1}^4 p_{ij}(E)|m_i><m_j|
\end{equation}
with variable $p_{ij}(E)$. The Poincare invariance of the elements $p_{ij}(\alpha)$ and $p_{ij}(E)$ strengthens the analogy between $\rho(\nu_\alpha)$ and $\rho(E)$. We shall call the mixed states $\rho(E)$ "dark neutrinos". All neutrino states engage in pure dephasing interactions and form the quantum environment. Quantum environment thus consists of neutrino environment and dark neutrino environment. All neutrino states have a mass 
\begin{equation}
m(\rho(E))=Tr(\hat{M} \rho(E))=\sum \limits_{i=1}^M p_{ii} m_i
\end{equation}
where $\hat{M}$ is the mass operator and they all engage in gravitational interactions.
  
It is our conjecture that the mixed dark neutrino states $\rho(E)$ form a distinct component of the dark matter. Today dark neutrinos are expected to be mostly non-relativistic cold particles. This suggests that hot dark neutrinos have been created in the early Universe and were redshifted to form the cold dark neutrinos environment at a later stage of the evolution. Hot dark neutrinos were still a part of dark matter at the times of galactic and large scale structure formation when most dark matter was cold or warm.

The interpretation of the quantum states $\rho(E)$ as dark neutrinos rests on the evidence for the light sterile neutrino. Unlike the dark neutrinos, the light sterile neutrinos $\nu_s$ mix with the active flavour neutrinos $\nu_e$, $\nu_\mu$ and $\nu_\tau$. The existence of the light sterile neutrino mixing with the active neutrinos was proposed~\cite{giunti07} to explain the observed anomalies in short-baseline neutrino oscillation experiments, namely the anomalous appearance of $\nu_e$ in $\nu_\mu \to \nu_e$ oscillations in accelerator experiments~\cite{LSND01,MiniBooNE10,MiniBooNE13} as due to $\nu_\mu \to \nu_s \to \nu_e$ and the anomalous disappearance of $\bar{\nu}_e$ and $\nu_e$ in Gallium~\cite{SAGE06,acero08}, reactor~\cite{acero08,mention11} and accelerator~\cite{T2K14} measurements as due to $\nu_e \to \nu_s$. The data require that the fourth mass eigenstate $|m_4>$ carries the heaviest mass $m_4 \approx 1$eV ~\cite{T2K14,palazzo13,conrad13,kopp13,giunti13}. The confirmation of the so far tentative  evidence for the existence of light sterile neutrino awaits a new generation of high precision accelerator, reactor and other short-baseline oscillations  experiments~\cite{conrad13,abazajian12,rubbia13,antonello13,kayser14,adey14,katori14,huber14} and in experiments using dark matter detectors~\cite{coloma14}. 

Independent evidence for light sterile neutrino comes also from cosmology. Planck satellite has exposed a tension between several early and late time observables in the minimal inflationary $\Lambda$CDM Model. In particular the Planck measurement~\cite{ade13} of the Hubble parameter $H_0$ is in tension with local measurements by Hubble Space Telescope (HST)~\cite{riess11,freedman12}. All these tensions can be resolved by adding a single light sterile neutrino to the standard $\Lambda$CDM  model~\cite{archidiacono13,hamann13,wyman14,battye14,zheng14,bergstrom14}. 

\subsection{Conversion of flavour neutrinos into dark neutrinos in pure dephasing interactions}

Propagation of free neutrinos is a unitary evolution. All unitary evolutions  evolve pure states into pure states and mixed states into mixed states. As a result there is no conversion of active neutrinos into the dark neutrinos, and vice versa, during the propagation of active neutrinos and dark neutrinos. Oscillations of active neutrinos are unitary transformations of the initial pure production state.
To explain why dark neutrinos are "cold" particles we first show that in dephasing interactions flavour neutrino states are converted into dark matter states. 

Consider a scattering process with the $S$-matrix final state
\begin{equation}
\rho_f(S)=\sum \limits_{\alpha,\beta} \sum \limits_{m,n} S_{\alpha,m}\rho(S_i)_{mn}S^*_{\beta,n}|\alpha><\beta|
\end{equation}
where $|\alpha>,|\beta>$ are final state vectors that include dipion or diparticle angular states $|J\lambda>$, $|m>,|n>$ are initial state vectors, $\rho(S_i)_{mn}$ is the initial state density matrix and $S_{\alpha,m}$ are $S$-matrix amplitudes. A single event of this scattering interacts with a single quantum state of the environment $\rho_i(E)$ given by (9.2). The interaction is unitary~\cite{svec13b} given by
\begin{equation}
\rho_f(S,E)=U\rho_f(S)\otimes \rho_i(E) U^+
\end{equation}
The joint density matrix has an explicit form
\begin{equation}
\rho_f(S,E)=\sum \limits_{\alpha',\beta'} \sum \limits_{i'j'}
\rho_f(S,E)_{\alpha'i',\beta'j'}|\alpha' m_{i'}><\beta' m_{j'}|
\end{equation}
where
\begin{equation}
\rho_f(S,E)_{\alpha'i',\beta'j'}=\sum \limits_{ij}\sum \limits_{\alpha,\beta}  p_{ij}(E_i) \rho_f(S)_{\alpha,\beta} <\alpha'i'|U|i \alpha ><\beta'j'|U|j \beta >^*
\end{equation}
To be in accord with the form (1.3) of the Kraus representation for $\rho_f(O)=Tr_E \rho_f(S,E)$ we must require~\cite{svec13b} 
$<\alpha' i'|U|i \alpha >= \delta_{i'i} <\alpha'i|U|i \alpha >$. With this constraint and taking a trace over the system $S$ we get the final state of the environment $\rho_f(E)$
\begin{eqnarray}
\rho_f(E) & = & Tr_S \rho_f(S,E)=\sum \limits_{ij} p_{ij}(E_i) a_{ij}(S) |m_i><m_j|\\
\nonumber
a_{ij}(S) & = & \sum \limits_{\alpha'}\sum \limits_{\alpha,\beta}
\rho_f(S)_{\alpha,\beta} <\alpha'i|U|i \alpha ><\alpha'j|U|j \beta >^*
\end{eqnarray}
Due to the unitarity of Kraus operators the parameters $a_{ii}(S)=1$. The diagonal terms $p_{ii}(E)$ are thus equal to the diagonal terms of the flavour neutrino state $p_{ii}(\alpha)=|U_{\alpha i}|^2$. It follows from (9.3) that dark neutrinos have the same mass $m_\alpha$ as the initial neutrino  $\nu_\alpha$ in the dephasing interaction. The probabilities $p_{ii}(\alpha)$ define the probabilities in the definition of bilinear terms of decohering amplitudes (4.4) and (4.5) for a single event of interaction of $\rho_f(S)$ with $\rho_i(E)$. In actual measurements the probabilities $p_{ii}(\alpha)$ are averaged in each $(m,t)$ bin over $N$ random events coming from both dark neutrinos and neutrino background. The random probabilities $p_{ii}(bin)=\sum \limits_\alpha n_\alpha p_{ii}(\alpha)$ where $n_\alpha=N_\alpha/N$ is the random probability of events with $p_{ii}(\alpha)$ given by their total number $N_\alpha$ for $\alpha=e,\mu,\tau,s$.

The terms $a_{ij}(S)$ depend on the scattering process and modify the elements $p_{ij}(E_i)$ of the initial state. In general, any pure initial state $\rho_i(E)$ will be converted into a mixed final state $\rho_f(E)$. Specifically, in dephasing interactions active neutrinos with $p_{ij}(\alpha)=U_{\alpha i}U^*_{\alpha j}$ will be converted into mixed states of dark neutrinos. Because dephasing interactions do not change the four-momenta of the particles~\cite{svec13b}, the produced dark neutrinos will carry the momentum $\vec{p}$ of the mass eigenstates of the initial active neutrinos. However note that since the momentum of the mass eigenstates $|m_k>$ is not an interacting degree of the environment the matrix elements of the Kraus operator $V_k$ and the parameters $a_{ij}$ do not depend on this momentum. The only change that takes place is the exchange of quantum information entropy between the states $\rho_f(S)$ and $\rho_i(E)$. For pure initial states the von Neumann entropy $S(\rho_i(E))=0$ while for the final mixed states $S(\rho_f(E))>0$.

\subsection{Genesis of relativistic dark neutrinos}

Pure neutrinos $\nu_e, \nu_\mu, \nu_\tau$ were created copiously in weak interactions of quarks, leptons and hadrons in the early stages of the Universe. Their oscillations would produce sterile neutrinos $\nu_s$. In dephasing interactions with a variety of particle scattering processes these pure neutrinos would be converted into the particles of hot dark neutrinos. Such dephasing processes would be essetially a continous generation of the dark matter for a period of time.

We consider quark-gluon plasma following the the electro-weak symmetry breaking at the energy scale $E\approx 1$ TeV corresponding to radiation temperature $T\approx 10^{16}$ K.
In addition to quarks and gluons the plasma includes photons, leptons  and active neutrinos. We shall focus first on four cosmological processes in this quark-gluon plasma
\begin{eqnarray}
\gamma + \ell & \to & \pi\pi + \ell, \quad \ell=e^{\pm},\mu^{\pm}, \tau^{\pm}\\
\gamma + q    & \to & \pi\pi + q, \quad q=u,d,s,\bar{u},\bar{d},\bar{s}\\
\ell^-+\ell^+ & \to & \pi\pi +\gamma\\
q + \bar{q}   & \to & \pi\pi + \gamma
\end{eqnarray}
Just like $\pi N \to \pi \pi N$ processes at low momentum transfers are produced largely by a pion exchange in the $t$-channel, so the processes (9.9) and (9.10) are produced by the virtual $\gamma$ exchange in the $t$-channel at low $t$. The produced pions dissolve quickly in the plasma up to the hadronization energy scale $\Lambda_{QCD}\approx 200$ MeV corresponding to the temperature $T_c \approx 10^{12.375}$ K~\cite{mukhanov05,bhalearo14,bazavov14}. Since hadron resonances have a very short life-time they decay rather than dissolve. At high energies the processes (9.9) and (9.10) have negligible cross-sections. We are interested in resonant production of the dipions at lower energies where the cross-sections are higher for dipion masses below 2000MeV. 

Since the leptons and quarks are thermalized we take their energy $E_b=m_b+kT$ and the photon energy $E_\gamma=kT$. The conservation of energy in (9.9) and (9.10) reads
\begin{equation}
\sqrt{s}=m_b+2kT=\sqrt{W^2+p^2}+\sqrt{m_b^2+p^2}
\end{equation}
where $p=|\vec{p}|$ is the centre-of-mass momentum and the dipion mass $W^2=(p_1+p_2)^2$ where $p_1$ and $p_2$ are the four-momenta of the two pions. The maximum value of the dipion mass $W$ at a given energy $kT$ occurs for $p=0$ which gives $W=2kT$. The resonances that couple to both $\gamma \gamma$ and $\pi\pi$ are $f_0(500)$, $f_0(980)$ and $f_0(1370)$ in the $S$-wave and $f_2(1270)$, $f_2'(1525)$ and $f_2(1950)$ in the $D$-wave. The strongest coupled resonance in the reactions (9.9) and (9.10) is $f_2(1270)$. To reach the $f_0(980)$ and $f_2(1270)$ the minimum temperature must be $10^{12.925}$ K ($kT=1,450$ MeV). To reach all three $D$-wave resonances the minimum temperature must be $10^{13.125}$ K ($kT=2,298$ MeV). 

The processes (9.11) and (9.12) are interactions similar to (9.9) and (9.10), respectively. They proceed by virtual photon exchange in the direct $s$-channel. In all four production processes the dipions are produced by virtual process
$\gamma\gamma(q^2)\to \pi\pi$ where $q^2$ is the four-momentum squared of the virtual photon. The process (9.11) is experimentally accessible and yields information on the differential cross-sections and total cross-section of $\gamma\gamma(q^2)\to \pi\pi$. Recently Belle Collaboration published high statistics measuremets of $\gamma\gamma(q^2)\to \pi^+\pi^-$~\cite{mori07a,mori07b} and $\gamma\gamma(q^2)\to \pi^0\pi^0$~\cite{uehara08,uehara09} for dipion masses $W=0.8 - 1.5$ GeV and $W=0.6-4.1$ GeV, respectively, extending previous lower statistics measurements~\cite{boyer90,behrend92,marsiske90}. Both reactions are dominated by $f_2(1270)$ resonance. The dipion mass $W=4.1$ GeV in the reaction (9.9) corresponds to the temperature $T= 10^{13.377}$ K well above the hadronization temperature $T_c$. In addition to the dipion states heavier dimeson states $M^+M^-$, $M^0\bar{M^0}$ and dibaryon states $B\bar{B}$ will be produced in the reactions (9.9)-(9.12) in the early Universe at $T>T_c$. Belle Collaboration recently published results for two-photon scattering at energies 2.4-4.1 GeV into $K^+K^-$~\cite{abe04,nakazawa05}, $K^0_S K^0_S$~\cite{chen07}, $D^+D^-$, $D^0\bar{D^0}$~\cite{uehara06,uehara08} and $p\bar{p}$~\cite{kuo05}. The colliding beams at Belle are unpolarized and thus do not allow amplitude analysis to observe spin mixing expected in the partial wave production amplitudes.

The differences between the unitary relative phases and the measured relative phases in $\pi^- p \to \pi^- \pi^+n$, $\pi^+ n \to \pi^+ \pi^-p$ and $\pi^- p \to \pi^0\pi^0 n$ indicated a pure dephasing final state interaction. These processes will convert flavour neutrinos into dark neutrinos but are prevented from doing so at and below the $\Lambda_{QCD}$ hadronization scale by the low energy $kT$. The spin formalism describing the $S$-matrix production of the final states $\rho_f(S)$ in the processes (9.9)-(9.12) is somewhat similar to that in $\pi N \to \pi \pi N$~\cite{svec13a}. While this formalism for the final state $\rho_f(S)$ is still to be developed the existence of the dephasing interactions of $\rho_f(S)$ is no longer contingent on the predictions, if any, of the unitary evolution law for these processes. Existence of such interactions has been established by the measurements of $\pi N \to \pi \pi N$ processes. We can surmiss that the processes (9.9)-(9.12) will also convert relativistic flavour neutrinos into the relativistic dark neutrinos. 

Dark neutrinos do not interact with particles of the Standard Model nor with any dark matter particles, and their dephasing interactions do not change their four-momentum. The dark neutrinos thus form a colissionless gas of distinguishable quantum states $\rho(E)$ that has characteristcs of relativistic dark matter. 

The energy density for the radiation dominated phase is given by~\cite{zuber12}
\begin{equation}
\rho_{rad}=\frac{\pi^2}{30}g_{eff}T^4
\end{equation}
where $g_{eff}$ represents the sum of all effectively contributing massless degrees of freedom
\begin{equation}
g_{eff}=\sum \limits_{\text{i=bosons}}g_i\Bigl(\frac{T_i}{T}\Bigr)^4 + \frac{7}{8}\sum \limits_{\text{i=fermions}}g_i\Bigl(\frac{T_i}{T}\Bigr)^4
\end{equation}
In this equation the equilibrium temperature $T_i$ of the particle $i$ is allowed to differ from the photon temperature $T$. The statistical weights are $g_\gamma=2$ for photons, $g_e=4$ for $e^-, e^+$ and $g_\nu=6$ for $\nu_\alpha=e,\mu,\tau$ assuming Dirac neutrinos. Since the produced dark neutrinos do not interact with any particles of the Standard Model, they immediately decouple trom the radiation environment. Even as they have relativistic momenta these fermions thus never contribute to $g_{eff}$ and to the radiation energy density $\rho_{rad}$. Instead they contribute a part to the energy density of the dark matter and their momenta decay as $p\sim a(t)^{-1}$ where $a(t)$ is the scale factor.

\subsection{Dark neutrinos and the formation of structure}

The clustering properties of the collisionless dark matter depend on the free streaming length $\lambda_{fs}$. At length scales larger than $\lambda_{fs}$ halos can form while at smaller length scales the formation of halos is damped~\cite{lesquorques13}. Today hot, warm, cool and cold components of dark matter have free streaming lengths of the order $\sim$ Mpc, $\sim$ kpc, $\sim $pc and $\sim$ mpc~\cite{boyanovsky08}.  

Due to the expansion of the space the relativistic dark neutrinos would be redshifted to become a thermal component of the dark matter today. We refer to this component as dark neutrino quantum environment. At the onset of galactic and large scale structure formation when the temperature of the Universe was about 1 keV dark neutrinos were still relativistic. The existence of cold or warm dark matter with the shorter free streaming lengths is required for the formation and stability of the galactic and large scale structures structures.
The free streaming length of dark neutrinos has not yet been precisely established but they are expected to account only for a part of dark matter because of their small mass.

A warm dark matter is expected to reconcile the large scale behaviour of the dark matter in $\Lambda$CDM model with the internal structure of the galaxies and with observed abundances of satellite galaxies. A possible candidate for warm dark matter is the sterile neutrino with mass 7.1 keV produced via lepton number-driven MSW resonant conversion of active neutrinos near or at the Big Bang Nucleosynthesis (BBN) epoch~\cite{shi99,abazajian02}. These massive sterile neutrinos are predicted to have a two-photon X-ray radiative decay at 3.55 keV~\cite{abazajian01,abazajian14}. A non-resonantly produced 7.1 keV sterile  neutrino~\cite{dodelson94,harada14} also predicts the 3.5 keV X-ray line. An emission line at 3.55-3.57 keV was recently detected in X-ray spectrum of galaxy clusters in two independent observations~\cite{bulbul14,boyarski14}.
These results suggest a multicomponent neutrino structure of the dark matter with dark neutrinos one such component. Dark neutrinos can form only a part of the dark matter but they still may have played a role in the structure formation and evolution. This is because the evolving free streaming length of dark neutrinos will depend not only on their neutrino mass $m_a$, $a=e,\mu,\tau,s$ but also on internal degrees of freedom related to their quantum states $\rho(E,a)$ such as the quantum entropy $S_a=S(\rho(E,a))$. 

Free streaming in a general mixed multicomponent dark matter was recently analysed by Boyanovsky~\cite{boyanovsky08}. The free streaming length today at $z=0$ of any Fermionic thermal relic as a unique component of dark matter reads~\cite{boyanovsky08}
\begin{equation}
\lambda_{fs,a}(0)=\frac{14}{g^{\frac{1}{3}}_{d,a} \sqrt{I_F[u]}}\Bigl(\frac{keV}{m_a}\Bigr) kpc
\end{equation}
where $g_{d,a}$ is the effective number of ultrarelativistic degrees of freedom at decoupling for each species $a$ of mass $m_a$, and the functions $I_F[u]$ are  dimensionless ratios of integrals of the distribution functions of the decoupled particles which are determined by the microphysics at the decoupling.

Consider now the case of active neutrinos or the light sterile neutrino $a=e,\mu,\tau,s$. For these particles the effective degree of freedom $g_{d,a}$ is small and their masses $m_a \ll$ keV. As the result their free streaming length will be large. The dark neutrinos $\rho(E,a)$ produced by flavour neutrino $\nu_a$ will have the same mass $m_a \ll$ keV but will be distinguishable by the values of their quantum entropy $S_a=S(\rho(E,a))$. These are the new discreet internal degrees of freedom of dark neutrinos. A large number of various dephasing interactions may produce the same value of $S_a$ but the total number $N_a$ of different discreet values of the entropy $S_a$ will be still very large. We identify $N_a$ with the effective number of degrees of freedom $g_{d,a}=N_a$. As the result the free streaming length of dark neutrinos will be substantially smaller than the free streaming length of active neutrinos.

From the data given in Ref.~\cite{boyanovsky08} we find 
$\frac{14} {\sqrt{I_F[u]}}=21.5443469$. This enables us to calculate the number of effective degrees of freedom $g_{d,a}=N_a$ assuming free streaming lengths $\lambda_{fs,a}(0)$ of hot, warm, cool and cold dark matter today for dark neutrino masses $m_e=1$ meV and $m_s=1$ eV. The results are summarized in the Table III.

\begin{table}
\caption{Effective internal numbers of freedom of dark neutrinos $\rho(E,e)$ and $\rho(E,s)$ identified as the number $N_e$ and $N_s$ of microstates of their quantum entropies $S_e$ and $S_s$, respectively, for assumed free streaming lengths of hot, warm, cool and cold dark matter.}
\begin{tabular}{|c|c|c|c|c|}
\toprule
$\text{dark matter}$ & $\text{hot}$ & $\text{warm}$ & $\text{cool}$ & $\text{cold}$\\
\colrule
$\lambda_{fs}(0)$ & 5 Mpc & 5 kpc & 5 pc & 5 mpc\\
\colrule
$N_e(m_e=1 meV)$ & $8\times10^{10}$ & $8\times10^{19}$ & $8\times10^{28}$ & $8\times10^{37}$\\
\colrule
$N_s(m_s=1 eV)$ & $80$ & $8\times10^{10}$ & $8\times10^{19}$ & $8\times10^{28}$\\
\botrule
\end{tabular}
\label{Table III.}
\end{table}
The values of $N_a$ represent the number of dephasing interactions of neutrinos $\nu_a$ which produced dark neutrinos $\rho(E,a)$ with differing values of the entropy $S_a$. Due to the large degeneracy of the values of $S_a$ the actual number of such dephasing interactions will be much larger.

The number of relic thermal neutrinos is estimated at $n_{\nu0}=336$      cm$^{-3}$~\cite{zuber12}. With the diameter of the observable Universe estimated at $93$ Glyr~\cite{halpern12}, the total number of the relic neutrinos in the Universe is $\sim 1.20\times 10^{89}$. We assume that the total number of dark neutrinos in the Universe is similar at $\sim 3.57\times10^{88}-3.57\times10^{89}$ corresponding to the number density of dark neutrinos $100-1000$ cm$^{-3}$. Assuming average degeneracy of the entropy microstates $\approx 10^{50} - 10^{52}$ the cold dark matter is favoured. Cool dark matter is favoured for higher average degeneracy $\approx 10^{60}-10^{61}$. An interesting possibility with such degeneracy is when $N_e=N_s=8\times10^{28}$ so that dark neutrinos $\rho(E,e)$ form cool dark matter while the dark neutrinos $\rho(E,s)$ form cold dark matter. These crude estimates suggest that dark neutrinos will be able to participate in the structure formation and evolution. The observable effects of dark neutrinos on this structure would be connected to the quantum information carried by the dark neutrinos. 

\subsection{Possible test of the dark neutrino hypothesis}

We identify the mixed neutrino states with particles of dark matter. In principle this hypothesis can be tested experimentally in dephasing interactions of pure neutrino states $\rho(E_i)=\rho(\nu_\alpha)$ with scattering processes $\pi N \to \pi \pi N$. In $\pi^- p \to \pi^- \pi^+ n$ the parameters $a_{ij}(S)$ in (9.8) are given by

\begin{eqnarray}
a_{ij}(S) & = & \sum \limits_{J\geq0,\lambda\leq1} \sum \limits_{\chi} \rho_f(S)^{J\frac{1}{2},J\frac{1}{2}}_{\lambda \chi,\lambda \chi}
|<J\lambda,\chi|V|J\lambda,\chi>|^2\\
\nonumber
& + & \sum \limits_{J\geq0,\lambda\leq1} \sum \limits_{\chi} 
2 Re \Bigl(\rho_f(S)^{J\frac{1}{2},J-1\frac{1}{2}}_{\lambda \chi,\lambda \chi} <J\lambda,\chi|V|J\lambda,\chi><J\lambda,\chi|V|J-1\lambda,\chi>^*\Bigr)\\
\nonumber
 & + & \sum \limits_{J,\lambda\geq2} \sum \limits_{\chi} 
\rho_f(S)^{J\frac{1}{2},J\frac{1}{2}}_{\lambda \chi,\lambda \chi}<J\lambda,\chi|V_i|J\lambda,\chi><J\lambda,\chi|V_j|J\lambda,\chi>^*\\
 & = & 1-\sum \limits_{J,\lambda\geq2} \sum \limits_{\chi} 
\rho_f(S)^{J\frac{1}{2},J\frac{1}{2}}_{\lambda \chi,\lambda \chi}
\Bigl(1-<J\lambda,\chi|V_i|J\lambda,\chi><J\lambda,\chi|V_j|J\lambda,\chi>^*
\Bigr)
\end{eqnarray}
In $\pi^- p \to \pi^0 \pi^0 n$ they are given by
\begin{eqnarray}
a_{ij}(S) & = & \sum \limits_{J,\lambda\geq0} \sum \limits_{\chi} 
\rho_f(S)^{J\frac{1}{2},J\frac{1}{2}}_{\lambda \chi,\lambda \chi}<J\lambda,\chi|V_i|J\lambda,\chi><J\lambda,\chi|V_j|J\lambda,\chi>^*
\end{eqnarray}
In (9.17) and (9.19) we have used a momentum projection of $\rho_f(S)$ given by (4.7) and constraints on matrix elements of Kraus operators given by (6.20) in Ref.~\cite{svec13b}. In (9.17) we used the orthogonality of matrix elements $<J \lambda,\chi|V_\ell|K\lambda,\chi>$ with non-zero elements for $K=J-1,J,J+1$. We assume that all amplitudes with $\lambda \leq1$ are decoherence free and that all amplitudes with $J,\lambda \geq2$ are decohering. $\chi$ is the recoil nucleon helicity. Due to the invariance of the trace of the density matrix $Tr\rho_f(S)=1$ under unitary transformations the expression (9.17) is equivalent to (9.18). The diagonal elements $a_{ii}(S)=1$ in (9.18) and (9.19). We see from (9.18) and (9.19) that the modifications of the initial non-diagonal elements $p_{ij}(E_i)$ are entirely due to the decohering amplitudes. For the pure state of neutrino $\nu_\alpha$ the initial elements are $p_{ij}(E_i)=p_{ij}(\alpha)=U_{\alpha i}U^*_{\alpha j}$.

The rates of dark matter generation in active neutrino interactions with $\pi N \to \pi \pi N$ final states will be largest in the resonant dipion mass regions. The production of dipion events is largest at low momentum transfers $|t|\leq 0.20$ (GeV/c)$^2$. However the decohering amplitudes $D^{2U},D^{2N}$ in $\pi^- p \to \pi^- \pi^+ n$ are larger only at higher $t$ and at higher dipion masses, which requires higher beam energy. Consider a pure hydrogen target exposed to a high intensity $\pi^-$ pion beam at energy optimized for maximum rate of event production in $\pi^- p \to \pi^- \pi^+ n$ and $\pi^- p \to \pi^0 \pi^0 n$  at $\rho^0(770)$ and at $f_2(1270)$ masses in a broad momentum transfer range. The target is exposed also to a stable high intensity $\nu_e$ neutrino beam. The neutrino flux through the target area is neasured over a period of time with the pion beam "on" and "off". When the pion beam is "off" there are no $\pi^- p \to \pi^- \pi^+ n$ and $\pi^- p \to \pi^0 \pi^0 n$ events to interact with the incident neutrinos. When the pion beam is "on" some of the $\pi^- p \to \pi^- \pi^+ n$ and $\pi^- p \to \pi^0 \pi^0 n$ events will interact with the pure neutrinos from the neutrino beam. These neutrinos become unmeasurable mixed states leading to a measurable drop in the neutrino flux akin to the disappearance anomaly. The experiment can prove the dephasing interaction of the flavour neutrinos and their conversion into the mixed neutrino states. The experiment is obviously technically difficult. However such experiments could advance laboratory evidence for particle structure of dark matter.

\section{Conclusions and outlook.}

We have presented the first amplitude onalysis of $\pi^- p \to \pi^- \pi^+ n$ production on polarized target using spin mixing mechanism to investigate spin mixing in decoherence free amplitudes and dephasing in decohering amplitudes. We find a unique solution for the $S$-and $P$-wave spin mixing and $S$-matrix amplitudes and for the spin mixing parameters in $\rho^0(770)-f_0(980)$ mixing. Our method extracts $D$-wave observables from the CERN data which allows a model dependent determination of the $D$-wave amplitudes. We find a clear $\rho^0(770)$ signal in the amplitudes $|D^U|^2$ and $|D^N|^2$ but cannot determine its spin mixing parameters. Decohering amplitudes do not mix spin and possess $M$ distict phases where $2\leq M \leq 4$. Our analysis of decohering  amplitude $D^{2U}$ determines that $M=4$. We can assume that all amplitudes with $J \geq 2$ and $\lambda \geq 2$ are decohering. 

We have used the spin mixing transversity amplitudes $S_\tau$ and $L_\tau$ to determine the spin mixing helicity amplitudes $S_n$ and $L_n$. Similarly, we have used the $S$-matrix transversity amplitudes $S^0_\tau$ and $L^0_\tau$ to determine the $S$-matrix helicity amplitudes $S^0_n$ and $L^0_n$. Using these helicity amplitudes we then determine the phase-shifts $\delta_P$ and $\delta^0_S$ for the spin mixing and for the $S$-matrix amplitudes below the $K\bar{K}$ threshold. The two Solutions for $\delta^0_S$ for the spin mixing amplitudes pass through $90^\circ$ which reflects the presence of $\rho^0(770)$ in the $S$-wave amplitudes. Apart from the Solution 1 in the mass region 830-930 MeV, these two Solutions are similar to the Cracow Solutions "down-flat" and "down-steep", respectively. In contrast, the two Solutions for the $S$-matrix amplitudes are essentially flat and show no sign of $\rho^0(770)-f_0(980)$ mixing. Their near equality suggests that a unique solution for the phase-shift $\delta^0_S$ consistent with the $S$-matrix unitarity is attainable from amplitude analyses of data on polarized targets.

We present a physically motivated model of the quantum states $\rho(E)$ as mixed states of neutrino mass eigenstates. We identify these dark neutrinos with particles of a distinct component of dark matter. What we call quantum environment consists of cold dark neutrinos and all flavour neutrinos. Hot dark neutrinos were created in continous dephasing interactions of flavour  neutrinos with a variety of particle scattering processes for a period of time in the early Universe at temperatures above hadronization temperature $\approx$ 200 MeV. Cooled by the expansion of the space this hot dark matter has evolved into a late cold component of cold dark matter. A careful analysis of the formation of galactic and large scale structures in the Universe indicates that most dark matter should be cold or warm at the onset of the  galaxy formation when the temperature of the Universe was about 1 keV. At these temperatures dark neutrinos were still relativistic and can account only for not yet specified part of the dark matter. However dark neutrinos may have participated in the structure formation and evolution because their free streaming length is shortened by the large number of effective degrees of freedom identified with the microstates of their quantum entropy. With an estimated present $\lambda_{fs}(0) \sim 5$ pc or $\sim 5$ mpc they can contribute to cool or cold dark matter or even to both. The observable effects of dark neutrinos on the galactic and large scale structure would be connected to the quantum information carried by the dark neutrinos. 
 
In priciple the model is experimentally testable. If confirmed what do we learn about dark neutrinos and dark matter from the measurements of $\pi N \to \pi \pi N$ processes?

The experimental information on decohering amplitudes and their phases would determine the parameters $a_{ij}(S)$ in (9.18) and (9.19). Together with the neutrino mixing matrix $U_{\alpha i}$ the momentum independent parameters $a_{ij}(S)$ would fully determine the quantum state $\rho_f(E)$ of the dark neutrinos produced in dephasing interactions of active neutrinos with $\pi N \to \pi \pi N$ processes. Thus neutrino oscillation experiments and pion production experiments alone could render a certain class of dark neutrino particles experimentally observable even though their momentum is not observable. However apart from the case of the amplitudes $D^{2U}_u$ and $D^{2N}_u$ discussed in the Section VI.D together with the approximation $D^{2U}_d=D^{2N}_d=0$ the phases of the decohering amplitudes in $\pi N \to \pi\pi N$ processes remain to be determined. The quantum states $\rho_f(E)$ of dark neutrinos produced in the dephasing interactions of the initial dark neutrinos $\rho_i(E)$ from the cold dark matter are not observable in $\pi N \to \pi\pi N$ processes rendering in turn unobservable the dark matter states $\rho_i(E)$.

The pure dephasing interactions involve only the exchange of quantum information entropy between the two states $\rho_f(S)$ and $\rho_i(E)$. As such they stand outside of the Standard Model. In $\pi^- p \to \pi^- \pi^+ n$ this exchange is associated with spontaneous violation of rotational/Lorentz symmetry in the spin mixing amplitudes $(S_\tau, L_\tau)$, $(U_\tau, D^U_\tau)$ and $(N_\tau, D^N_\tau)$~\cite{svec13b}. The measurements of the spin mixing parameters and of the phases of decohering amplitudes provide information about the dynamics of the dephasing interactions of the dark matter.

The spin formalism developed in Ref.\cite{svec13a} and in this work applies equally well to a number of other meson production processes such as $K N \to K \pi N$, $\pi N \to \pi K \Lambda$, $\overline{K} N \to \pi \pi \Lambda$ and others. These processes could be measured on polarized target and the measurements with $\Lambda$ would also allow measurements of recoil $\Lambda$ polarization by its weak decays. Modern polarized targets reach high values of polarization and enable to select an arbitrary direction of the polarization vector~\cite{leader01}. The non-standard interactions of particle scattering processes with the quantum environment of thermal dark neutrinos and active neutrinos background are not rare events although they require high statistics experiments with polarized targets to detect their signatures: spin mixing and the violation of cosine conditions. The presented amplitude analysis illustrates this new kind of search for dark matter.


\end{document}